\documentclass[12 pt]{article}
\usepackage{amsmath,amssymb}
\usepackage[arc,ps,dvips,curve]{xy}
\usepackage{epsf}
\usepackage[]{amssymb, amsthm, graphics, color}
\usepackage{amsmath,amssymb,amsthm}
\usepackage[pdftex]{graphicx}
\usepackage[dvipsnames]{xcolor}
\usepackage{hyperref}

\UseDVIPSspecials\relax
\def\connsumtransversevertex#1#2#3{\setbox0=\hbox{$\scriptstyle\scriptstyle\rightarrow$}#1\#_{\vtop{\hbox to \wd0{\hfil$\scriptstyle #3$\hfil}\kern-7pt\hbox{$\scriptstyle\rightarrow$}}} #2}

\def\set#1\endset{\{\,#1\,\}}
\def\proof{\ifdim\lastskip<\smallskipamount\relax\removelastskip
  \vskip\smallskipamount\fi\leavevmode\hbox to 10pt{\hfil}{\it Proof. }}
\def\strutdepth{\dp\strutbox}
\def\epmarker{\vbox to \strutdepth{\baselineskip\strutdepth\vss\hfill{%
\hbox to 0pt{\hss\vrule height 4pt width 4pt depth 0pt}\null}}}
\def\edproofmarker{\strut\vadjust{\kern-2\strutdepth\epmarker}}
\def\endproof{\edproofmarker\vskip10pt}
\def\table#1\endtable{\begin{tabular}#1 \end{tabular}}
\def\smallbullet{\raise.5pt\hbox{\scriptsize$\,\bullet\,$}}

\date{}

\begin{document}

\newtheorem{corollary}{Corollary}[section]
\newtheorem{theorem}{Theorem}[section]
\newtheorem{lemma}{Lemma}[section]

\newtheorem{proposition}{Proposition}[section]
\newtheorem{definition}{Definition}[section]

\title{Determining satisfiability of  $3$-SAT in polynomial time}
\author{ Ortho Flint, Asanka Wickramasinghe, Jay Brasse, Chris Fowler}
\maketitle

\vskip 2mm

\begin{abstract}
In this paper, we provide a polynomial time (and space), algorithm that determines satisfiability of  $3$-SAT.  The complexity analysis for the algorithm takes into account no efficiency and yet provides a low enough bound, that efficient versions are practical with respect to today's hardware. We accompany this paper with a serial version of the algorithm without non-trivial  efficiencies (link: polynomial3sat.org).

\end{abstract}

\section{Introduction}
 \noindent The Boolean satisfiability problem (or SAT), was the first problem to be proven  NP-complete by Cook in $1971$  \cite{cook71}.
$3$-SAT is one of Karp's original $21$ NP-complete problems \cite{karp72}.  At last count, there are over $3,000$ NP-complete problems considered to be important. Any problem in NP can be presented as a SAT problem.
A SAT problem can be transformed into a $3$-SAT problem, where any increase in the number of clauses is a polynomial bound.
 We provide a deterministic polynomial time algorithm that decides satisfiability of  $3$-SAT. We also provide in Big O notation the complexity of the algorithm without efficiencies. The original serial version is called $\it{naive}$, by which we mean, none of the efficiencies that could be used for dramatic reductions in work are present. However, this version is quite useful for research, as we can attest.   There are many non-trivial efficiencies that can be incorporated. \\

\noindent Note well, that the intent of an informal presentation of the algorithm, is to offer people that are outside the math/computer science community, a fairly straightforward explanation. \\

\section{Preliminaries and Definitions}

\begin{definition}\label{definition: $3$-SAT}

A $3$-SAT is a collection of literals or variables (usually represented by integers), in groups called clauses where no clause has more than $3$ literals, and at least one clause does have $3$ literals. If it's possible to select exactly one literal from each clause such that no literal\hspace{.05cm} $l$, and its negation $-l$ (or  denoted $ \neg{\hspace{.05cm}l}$, meaning not $l$),  appear in the collection of chosen literals, we say that the $3$-SAT is satisfiable, otherwise we say it's  unsatisfiable. For satisfiability, the collection of literals chosen is called a solution, which is also known as a truth assignment. Note that the size of a solution set is smaller than the size of the collection, if the collection had at least two clauses of which the same literal was chosen.

\end{definition}

\noindent It is important to note, that our definition of a solution (definition $2.12$), is inextricably tied to our constructs for a collection of clauses.\\

\noindent When we think of literals (also called $\textit{atoms}$), we can consider an edge joining two vertices, each with an associated literal, if and only if, it's not a literal and its negation. Under no circumstance would a literal and its negation be connected by an edge. There are also no edges between two literals from the same clause. So conceptually, there exist edges that connect every literal to every other literal with the two restrictions that were just stated. Then it follows, that a collection of literals for some solution is such that, a literal from each clause is connected to every other literal from that collection. In graph theory, such a graph is called a $\textit{complete}$ graph $K_{n}$, $n$ being the number of vertices, which here it's also the number of clauses. We shall denote this graph as: $K_{C}$ where $c$ is the number of clauses.\\

\begin{definition}\label{definition: edge-sequence} An edge-sequence is an ordered sequence with elements $1$ and $0$. The ordering is an ordering of the clauses, with indexing: $C_{1}$, $C_{2}$, $C_{3}$, \hspace{.025cm}$\ldots$ \hspace{.025cm},  $C_{c}$ where a corresponding $C_{i}$ has its literals ordered the same way for each sequence constructed for a $3$-SAT. An edge-sequence $I$, for an edge with endpoints labelled $x$ and $y$, where $x$ $\ne$ $-y$, the literals associated with the endpoints, is denoted by $I_{x,  y}$. The endpoints must always be from different clauses. We call the positions in $I_{x,  y}$ that correspond to a clause $C_{i}$ the cell $C_{i}$. The cells  $C_{j}$ and $C_{k}$ containing the endpoints, $x$ and $y$ for $I_{x,  y}$ have only one entry that is $1$ in the positions associated to $x$ and $y$.  When an edge-sequence is constructed, a given position in $I_{x,  y}$ is $1$ if the associated literal is not $-x$ or $-y$. The initial construction of $I_{x,  y}$ is subject to certain rules defined in $2.8$ and $2.9$, which may produce more zero entries. Lastly, removing one or more cells from $I_{x,  y}$  is again a (sub) edge-sequence, denoted by $I_{x, y}$*,  if the cells containing the endpoints for  $I_{x,  y}$ remain.

\end{definition}

\vskip 2mm

\begin{definition}\label{definition: edge-sequence pure literal}
We call a literal $x$, an edge-pure literal, if its negation does not appear in an edge-sequence. ie. there are zero entries in the positions associated with literal $-x$ in the edge-sequence, or no $-x$ exists. Note that the literals associated to the endpoints are necessarily edge-pure literals.

\end{definition}

\vskip 2mm

\begin{definition}\label{definition: singleton}
We call a literal $x$, an edge-singleton, if it has only one position in an edge-sequence. We say that only one position exists in each endpoint cell which correspond with the endpoints. And we call a literal a singleton, if it appears in only one clause for a  given collection of clauses.

\end{definition}

\vskip 2mm

\begin{definition}\label{definition: loner cell}
 A loner cell contains just one $1$ entry for some literal. And a loner clause contains just one literal.

\end{definition}

\vskip 2mm

\begin{definition}\label{definition: vertex-sequence}
A vertex-sequence is an ordered sequence with elements $1$ and $0$. \hspace{.025cm}The ordering is an ordering of the clauses, with indexing: $C_{1}$, $C_{2}$, $C_{3}$, \hspace{.025cm}$\ldots$ \hspace{.025cm}, $C_{c}$ where a corresponding $C_{i}$ has its literals ordered the same way for each sequence constructed for a $3$-SAT. A vertex-sequence $V$, for a vertex associated with literal $x$, is denoted by $V_{x}$. We call the positions in $V_{x}$ that correspond to a clause $C_{i}$ the cell $C_{i}$. The cell  $C_{j}$ containing the vertex $x$ for $V_{x}$ has only one entry that is $1$ in the position associated to $x$.  When a vertex-sequence is constructed, a given position in $V_{x}$ is $1$ if the associated literal is not $-x$. The initial construction of $V_{x}$ is subject to certain rules defined in $2.8$ and $2.9$,  which may produce more zero entries. Removing one or more cells from $V_{x}$  is again a (sub) vertex-sequence, denoted by $V_{x}$*,  if the cell containing $x$ remains.

\end{definition}

\begin{definition}\label{definition: bit-change and refinement}
When an entry $1$ in an edge-sequence or a vertex-sequence, becomes zero, we call it a bit-change. If a bit-change has occurred in an  edge-sequence or a vertex-sequence, we say the sequence has been refined, or a refinement has occurred. A zero entry never becomes a $1$ entry.

\end{definition}

\noindent It's worth noting here that if an edge-sequence $I_{a,b}$  has a zero entry in some position for a literal $c$, then there is no $K_{C}$, using literals $a, b$ and $c$ together. In fact, this is what a bit-change is documenting in an edge-sequence.  \\

\begin{definition}\label{definition: LCR rule}
The loner cell rule, LCR, is that no negation of a literal belonging to a loner cell can exist in an edge-sequence or vertex-sequence.
If such a scenario exists in an edge-sequence $I_{x,  y}$  or a vertex-sequence $V_{x}$  where $z$ is the loner cell literal, then all positions associated with literal $-z$ incur a bit-change. If this action of a bit-change for $-z$, creates another loner cell where the negation of the literal in the newly created loner cell is still present in  $I_{x,  y}$ or $V_{x}$  the action of a bit-change for the negation is repeated. Hence, to be LCR compliant may be recursive, but all refinements are permanent for any edge or vertex sequence.\\

\noindent LCR compliancy is determined for an edge-sequence $I_{x,  y}$  or a vertex-sequence $V_{x}$  if either $I_{x,  y}$ or $V_{x}$  is being constructed.
LCR compliancy is determined after any intersection  between two of more edge-sequences is performed.
LCR compliancy is determined if  an edge-sequence or vertex-sequence incurred any refinement.
\end{definition}

\begin{definition}\label{definition: K-rule }
The $K$-rule is that no cell from an edge-sequence $I_{x,  y}$  or a vertex-sequence $V_{x}$  can have all zero entries.
If this is the case, then $I_{x,  y}$ or $V_{x}$, equals zero, and $I_{x,  y}$ is removed from its $S$-set, or  $V_{x}$ is removed from the vertex-sequence table. Note that their respective removals, is a refinement.\\

\noindent  $K$-rule compliancy is determined for an edge-sequence $I_{x,  y}$ or a vertex-sequence $V_{x}$, if either $I_{x,  y}$ or $V_{x}$  are being constructed.
$K$-rule compliancy is determined after any intersection  between two of more edge-sequences is performed.
 $K$-rule compliancy is determined if  an edge-sequence or vertex-sequence incurred any refinement.
And finally, the $K$-rule is violated if all the vertex-sequences associated with a clause, are zero. In such a case, it's reported that the $3$-SAT is unsatisfiable.

\end{definition}

\noindent Before we work through an example, we must define what it means to take an intersection or union of two or more edge-sequences. No intersections or unions are taken with vertex-sequences.\\

\begin{definition}\label{definition: intersections and unions}

We take the intersection or union of two $n$ length edge-sequences, $A$ and $B$,  by comparing  position $i$ of $A$ and $B$, using the Boolean rules for  $\textit{intersections}$ (denoted by $\cap$), and $\textit{unions}$ (denoted by $\cup$), for all positions, $i$  $=$ $ 0, 1, 2, \ldots, n$$-1$.\\

\noindent Recall that the entry for position $i$ of $A$ and $B$, is either $1$ or $0$.\\

\noindent Then for an intersection, we have: \\

\noindent $1_{A}$ $\cap$ $0_{B}$ = $0_{A}$ $\cap$ $1_{B}$ = $0_{A}$ $\cap$ $0_{B}$ = $0$. And $1_{A}$ $\cap$ $1_{B}$ = $1$. \\

\noindent And for a union we have: \\

\noindent $1_{A}$ $\cup$ $0_{B}$ = $0_{A}$ $\cup$ $1_{B}$ = $1_{A}$ $\cup$ $1_{B}$ = $1$. And $0_{A}$ $\cup$ $0_{B}$ = $0$.\\

\end{definition}

\noindent In general, an intersection or union between two or more edge-sequences, is a multi-edged sequence. There is one exception described in section $3$, where an intersection or a union of intersections of edge-sequences, always produces an edge-sequence. \\

\begin{definition}\label{definition: S-set}
 An $S$-set is a collection of edge-sequences whose endpoints are from two clauses, $C_{i}$ and $C_{j}$ where $i \ne j$. The number of constructed edge-sequences to be an  $S$-set is $|C_{i}| | C_{j}|$ minus the non edge-sequences of the form: $I_{l,-l}$.   For $3$-SAT, there can be at most 9 edge-sequences in an $S$-set.
\end{definition}

\vskip 2mm

\begin{definition}\label{definition: a solution}

 A solution for a collection of $c$ clauses must have a corresponding collection of edge-sequences, for some $K_{C}$. More precisely, the intersection of all the edge-sequences together, for a $K_{C}$,  does not equal zero. ie. A solution $K_{C}$ exists if  \hspace{.05cm}$\underset{i, j}{\bigcap}\hspace{.05cm} I_{i, j}$ $\ne$ $0$, where $i$ and $j$ are every pair of endpoints from the collection of edges for a $K_{C}$.  A  $K_{P}$, $p<c$, exists if  the set of all sub edge-sequences $\mathcal{D}$  for $K_{P}$  are such that the intersection of  $\mathcal{D}$ does not equal zero. ie. A $K_{P}$ exists if  \hspace{.05cm}$\underset{i, j}{\bigcap}\hspace{.05cm} I_{i, j}$* $\ne$ $0$,   where $i$ and $j$ are every pair of endpoints from the collection of edges for  $K_{P}$. It is to be understood that an edge-sequence for a $K_{C}$ or  $K_{P}$, means the edge-sequence associated with an edge for a $K_{C}$ or  $K_{P}$.

 \end{definition}

\begin{center}
\large { \textbf{Example}}

\end{center}

\begin{center}
\includegraphics[width=5in,height=5in]{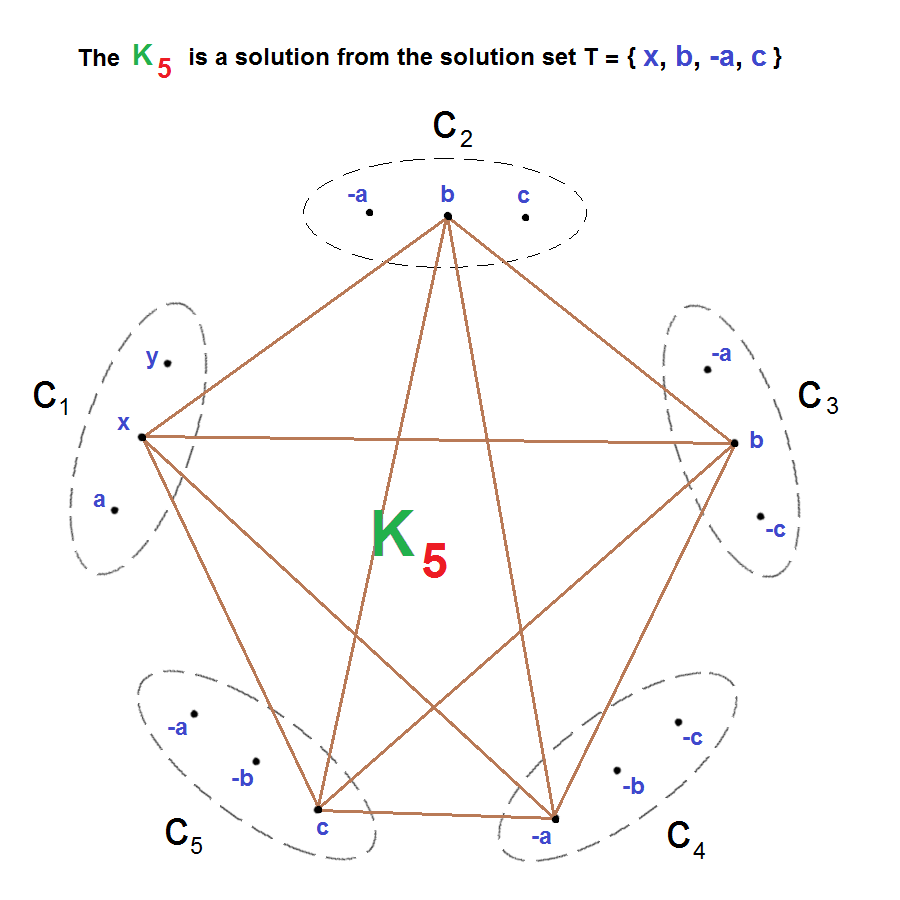}
\end{center}
 \begin{center}
 Figure $1$: A $3$-SAT with $5$ clauses
 \end{center}

\noindent We believe it is useful to think of $3$-SATs in terms of their corresponding graphs. For example, Figure $1$ (the only figure in the paper), depicts a $3$-SAT with five clauses. The clauses are numbered and each clause has three literals. We say that a  clause with three literals, thus three vertices, is $\textit{size}$ $3$. For $3$-SAT, clauses can only be of size $1$, $2$ or $3$. Although there would be an edge between any two literals not belonging to the same clause and not between a literal and its negation, we have chosen to show in  Figure $1$, just the edges for a $K_{5}$ from the solution set $T$ = $\{x, b, -a, c \}$. \\

\newpage

\noindent Of course, the challenge when presented with a $3$-SAT is to find at least one $K_{C}$, when it's a proper subgraph, or to determine that no $K_{C}$ exists. \\

\vskip 2mm
\noindent This example is to demonstrate the construction of edge-sequences for an $S$-set. We also provide one example of an intersection and  another of a union between two edge-sequences.\\

\vskip 2mm
\noindent For an $S$-set with edge-sequences formed using clauses $C_{i}, C_{j}$, we write: $S_{i, j}$. Recall, that an edge's endpoints are always labelled by their associated literals. Then the edge-sequence $I$, for an edge with endpoints with labelling $a, b$, is denoted by $I_{a, b}$. Below, we add sub-subscripts for our example (Figure $1$), $3$-SAT's edge-sequences, to indicate which clauses the endpoints are from. And within the (ordered) sequences, the subscripts indicate which literal received a zero or a one.\\

\vskip 2mm
\noindent Our example has $10$ $S$-sets and at most $9$ edge-sequences in each. With respect to the sub-subscripts, all the edge-sequences are unique. We will show just the initial $8$ edge-sequences (there is no  $I_{a_{1}, -a_{2}}$, a literal and its negation), for  $S_{1,2}$. We will also construct $I_{b_{3}, -c_{4}}$ from $S_{3,4}$ for the purpose of showing  intersections  and unions  of edge-sequences. \\

\vskip 2mm

\noindent Then the constructed edge-sequences for $S_{1,2}$, before $LCR$ and $K$-rule determination is:\\

\noindent \hspace{1.6cm} $ \overbrace{}^{C_{1}}$ \hspace{.9cm} $ \overbrace{}^{C_{2}}$ \hspace{1.12cm}  $ \overbrace{}^{C_{3}}$ \hspace{1.45cm} $ \overbrace{}^{C_{4}}$ \hspace{1.5cm} $ \overbrace{}^{C_{5}}$

\vskip 2mm
\noindent $I_{a_{1}, b_{2}} : ( 1_{a}, 0_{x}, 0_{y} \hspace{.15cm}\textcolor{brown}{|} \hspace{.15cm}0_{-a}, 1_{b}, 0_{c} \hspace{.15cm}\textcolor{brown}{|} \hspace{.15cm}  0_{-a}, 1_{b}, 1_{-c} \hspace{.15cm}\textcolor{brown}{|} \hspace{.15cm} 0_{-a}, 0_{-b}, \textcolor{Green}{1}_{-c}    \hspace{.15cm}\textcolor{brown}{|} \hspace{.15cm}   0_{-a}, 0_{-b}, \textcolor{red}{1}_{c} )$\\

\vskip 2mm
\noindent $I_{a_{1}, c_{2}} : ( 1_{a}, 0_{x}, 0_{y} \hspace{.15cm}\textcolor{brown}{|} \hspace{.15cm}0_{-a}, 0_{b}, 1_{c} \hspace{.15cm}\textcolor{brown}{|} \hspace{.15cm}  0_{-a}, \textcolor{Green}{1}_{b}, 0_{-c} \hspace{.15cm}\textcolor{brown}{|} \hspace{.15cm} 0_{-a}, \textcolor{red}{1}_{-b}, 0_{-c}    \hspace{.15cm}\textcolor{brown}{|} \hspace{.15cm}   0_{-a}, 1_{-b}, 1_{c} )$\\

\vskip 2mm
\noindent \hspace{-.4cm} $I_{x_{1}, -a_{2}} :( 0_{a}, 1_{x}, 0_{y} \hspace{.15cm}\textcolor{brown}{|} \hspace{.15cm}1_{-a}, 0_{b}, 0_{c} \hspace{.15cm}\textcolor{brown}{|} \hspace{.15cm}  1_{-a}, 1_{b}, 1_{-c} \hspace{.15cm}\textcolor{brown}{|} \hspace{.15cm} 1_{-a}, 1_{-b}, 1_{-c}    \hspace{.15cm}\textcolor{brown}{|} \hspace{.15cm}   1_{-a}, 1_{-b}, 1_{c} )$\\

\vskip 2mm
\noindent $I_{x_{1}, b_{2}} : ( 0_{a}, 1_{x}, 0_{y} \hspace{.15cm}\textcolor{brown}{|} \hspace{.15cm}0_{-a}, 1_{b}, 0_{c} \hspace{.15cm}\textcolor{brown}{|} \hspace{.15cm}  1_{-a}, 1_{b}, 1_{-c} \hspace{.15cm}\textcolor{brown}{|} \hspace{.15cm} 1_{-a}, 0_{-b}, 1_{-c}    \hspace{.15cm}\textcolor{brown}{|} \hspace{.15cm}   1_{-a}, 0_{-b}, 1_{c} )$\\

\vskip 2mm
\noindent $I_{x_{1}, c_{2}} : ( 0_{a}, 1_{x}, 0_{y} \hspace{.15cm}\textcolor{brown}{|} \hspace{.15cm}0_{-a}, 0_{b}, 1_{c} \hspace{.15cm}\textcolor{brown}{|} \hspace{.15cm}  1_{-a}, 1_{b}, 0_{-c} \hspace{.15cm}\textcolor{brown}{|} \hspace{.15cm} 1_{-a}, 1_{-b}, 0_{-c}    \hspace{.15cm}\textcolor{brown}{|} \hspace{.15cm}   1_{-a}, 1_{-b}, 1_{c} )$\\

\vskip 2mm
\noindent \hspace{-.4cm} $I_{y_{1}, -a_{2}} : ( 0_{a}, 0_{x}, 1_{y} \hspace{.15cm}\textcolor{brown}{|} \hspace{.15cm}1_{-a}, 0_{b}, 0_{c} \hspace{.15cm}\textcolor{brown}{|} \hspace{.15cm}  1_{-a}, 1_{b}, 1_{-c} \hspace{.15cm}\textcolor{brown}{|} \hspace{.15cm} 1_{-a}, 1_{-b}, 1_{-c}    \hspace{.15cm}\textcolor{brown}{|} \hspace{.15cm}   1_{-a}, 1_{-b}, 1_{c} )$\\

\vskip 2mm
\noindent $I_{y_{1}, b_{2}} : ( 0_{a}, 0_{x}, 1_{y} \hspace{.15cm}\textcolor{brown}{|} \hspace{.15cm}0_{-a}, 1_{b}, 0_{c} \hspace{.15cm}\textcolor{brown}{|} \hspace{.15cm}  1_{-a}, 1_{b}, 1_{-c} \hspace{.15cm}\textcolor{brown}{|} \hspace{.15cm} 1_{-a}, 0_{-b}, 1_{-c}    \hspace{.15cm}\textcolor{brown}{|} \hspace{.15cm}   1_{-a}, 0_{-b}, 1_{c} )$\\

\vskip 2mm
\noindent $I_{y_{1}, c_{2}} : ( 0_{a}, 0_{x}, 1_{y} \hspace{.15cm}\textcolor{brown}{|} \hspace{.15cm}0_{-a}, 0_{b}, 1_{c} \hspace{.15cm}\textcolor{brown}{|} \hspace{.15cm}  1_{-a}, 1_{b}, 0_{-c} \hspace{.15cm}\textcolor{brown}{|} \hspace{.15cm} 1_{-a}, 1_{-b}, 0_{-c}    \hspace{.15cm}\textcolor{brown}{|} \hspace{.15cm}   1_{-a}, 1_{-b}, 1_{c} )$\\

\vskip 2mm

\noindent Observe below, that after $LCR$ and $K$-rule is applied, edges $I_{a_{1}, b_{2}}$ and $I_{a_{1}, c_{2}}$ will be zero. Since the loner cells $C_{4}$ and $C_{5}$ in  $I_{a_{1}, b_{2}}$ negate each other, applying $LCR$ to either cell (we apply to first encountered), causes the other cell to be an empty cell which violates the $K$-rule. Similarly,  the loner cells $C_{3}$ and $C_{4}$ in  $I_{a_{1}, c_{2}}$ negate each other, producing the same outcome.\\

\noindent Then applying $LCR$ to  $I_{a_{1}, b_{2}}$,   we have:\\

\noindent \hspace{8.1cm}  $ \overbrace{}^{LCR}$ \hspace{1.25cm} $ \overbrace{}^{All \hspace{.05cm}Zeroes}$

\vskip 2mm
\noindent $I_{a_{1}, b_{2}} : ( 1_{a}, 0_{x}, 0_{y} \hspace{.15cm}\textcolor{brown}{|} \hspace{.15cm}0_{-a}, 1_{b}, 0_{c} \hspace{.15cm}\textcolor{brown}{|} \hspace{.15cm}  0_{-a}, 1_{b}, 1_{-c} \hspace{.15cm}\textcolor{brown}{|} \hspace{.15cm} 0_{-a}, 0_{-b}, \textcolor{Green}{1}_{-c}    \hspace{.15cm}\textcolor{brown}{|} \hspace{.15cm}   0_{-a}, 0_{-b}, \textcolor{red}{0}_{c} )$\\

\vskip 2mm
\noindent Now, we apply $K$-rule to  $I_{a_{1}, b_{2}}$, thereby $I_{a_{1}, b_{2}}$ $=$ $0$:\\

\vskip 2mm
\noindent $I_{a_{1}, b_{2}} : ( 1_{a}, 0_{x}, 0_{y} \hspace{.15cm}\textcolor{brown}{|} \hspace{.15cm}0_{-a}, 1_{b}, 0_{c} \hspace{.15cm}\textcolor{brown}{|} \hspace{.15cm}  0_{-a}, 1_{b}, 1_{-c} \hspace{.15cm}\textcolor{brown}{|} \hspace{.15cm} 0_{-a}, 0_{-b}, \textcolor{Green}{1}_{-c}    \hspace{.15cm}\textcolor{brown}{|} \hspace{.15cm}   \textcolor{red}{0}_{-a}, \textcolor{red}{0}_{-b}, \textcolor{red}{0}_{c} )$\\

\vskip 2mm

\noindent Applying $LCR$ to  $I_{a_{1}, c_{2}}$, we have:\\

\noindent \hspace{5.675cm}  $ \overbrace{}^{LCR}$ \hspace{1.2cm} $ \overbrace{}^{All \hspace{.05cm}Zeroes}$

\vskip 2mm
\noindent $I_{a_{1}, c_{2}} : ( 1_{a}, 0_{x}, 0_{y} \hspace{.15cm}\textcolor{brown}{|} \hspace{.15cm}0_{-a}, 0_{b}, 1_{c} \hspace{.15cm}\textcolor{brown}{|} \hspace{.15cm}  0_{-a}, \textcolor{Green}{1}_{b}, 0_{-c} \hspace{.15cm}\textcolor{brown}{|} \hspace{.15cm} 0_{-a}, \textcolor{red}{0}_{-b}, 0_{-c}    \hspace{.15cm}\textcolor{brown}{|} \hspace{.15cm}   0_{-a}, 1_{-b}, 1_{c} )$\\

\vskip 2mm
\noindent Now, we apply $K$-rule to  $I_{a_{1}, c_{2}}$, thereby $I_{a_{1}, c_{2}}$ $=$ $0$:\\

\vskip 2mm
\noindent $I_{a_{1}, c_{2}} : ( 1_{a}, 0_{x}, 0_{y} \hspace{.15cm}\textcolor{brown}{|} \hspace{.15cm}0_{-a}, 0_{b}, 1_{c} \hspace{.15cm}\textcolor{brown}{|} \hspace{.15cm}  0_{-a}, \textcolor{Green}{1}_{b}, 0_{-c} \hspace{.15cm}\textcolor{brown}{|} \hspace{.15cm} \textcolor{red}{0}_{-a}, \textcolor{red}{0}_{-b}, \textcolor{red}{0}_{-c}    \hspace{.15cm}\textcolor{brown}{|} \hspace{.15cm}   0_{-a}, 1_{-b}, 1_{c} )$\\

\newpage

\noindent The other $6$ edge-sequences of $S_{1, 2}$ listed above, are $LCR$ and $K$-rule compliant. Thus, the construction of the edge-sequences resulted in only $6$ edge-sequences for $S_{1, 2}$. We  note here that if an $S$-set had no edge-sequences after construction where $LCR$ and $K$-rule was applied, it would mean there is no solution for the collection of clauses given. We show a $K_{5}$ in Figure $1$, so all $10$ $S$-sets must exist, because each $S$-set has an edge-sequence for a $K_{5}$.\\

\vskip 2mm

\noindent Below, we show the result of doing $I_{x_{1}, c_{2}} \cap I_{b_{3}, -c_{4}}$ and $I_{x_{1}, c_{2}} \cup I_{b_{3}, -c_{4}}$.\\

\vskip 2mm

\noindent The $LCR$ and $K$-rule compliant edge-sequence $I_{b_{3}, -c_{4}}$ of $S_{3,4}$ is:\\
\vskip 2mm

\noindent  \hspace{-.4cm} $I_{b_{3}, -c_{4}} : ( \textcolor{red}{0}_{a}, 1_{x}, 1_{y} \hspace{.15cm}\textcolor{brown}{|} \hspace{.15cm}1_{-a}, 1_{b}, 0_{c} \hspace{.15cm}\textcolor{brown}{|} \hspace{.15cm}  0_{-a}, 1_{b}, 0_{-c} \hspace{.15cm}\textcolor{brown}{|} \hspace{.15cm} 0_{-a}, 0_{-b}, 1_{-c}    \hspace{.15cm}\textcolor{brown}{|} \hspace{.15cm}   \textcolor{green}{1}_{-a}, 0_{-b}, 0_{c} )$\\

\vskip 2mm
\noindent And the $LCR$ and $K$-rule compliant edge-sequence $I_{x_{1}, c_{2}}$ is:\\

\vskip 2mm
\noindent $I_{x_{1}, c_{2}} : ( 0_{a}, 1_{x}, 0_{y} \hspace{.15cm}\textcolor{brown}{|} \hspace{.15cm}0_{-a}, 0_{b}, 1_{c} \hspace{.15cm}\textcolor{brown}{|} \hspace{.15cm}  1_{-a}, 1_{b}, 0_{-c} \hspace{.15cm}\textcolor{brown}{|} \hspace{.15cm} 1_{-a}, 1_{-b}, 0_{-c}    \hspace{.15cm}\textcolor{brown}{|} \hspace{.15cm}   1_{-a}, 1_{-b}, 1_{c} )$\\

\vskip 2mm

\noindent The intersection, $I_{x_{1}, c_{2}} \cap I_{b_{3}, -c_{4}}$ is :

\vskip 4mm

\noindent  \hspace{1.1cm} $( 0_{a}, 1_{x}, 0_{y} \hspace{.15cm}\textcolor{brown}{|} \hspace{.15cm}\textcolor{red}{0}_{-a}, \textcolor{red}{0}_{b}, \textcolor{red}{0}_{c} \hspace{.15cm}\textcolor{brown}{|} \hspace{.15cm}  0_{-a}, 1_{b}, 0_{-c} \hspace{.15cm}\textcolor{brown}{|} \hspace{.15cm} \textcolor{red}{0}_{-a}, \textcolor{red}{0}_{-b}, \textcolor{red}{0}_{-c}    \hspace{.15cm}\textcolor{brown}{|} \hspace{.15cm}   1_{-a}, 0_{-b}, 0_{c} )$\\

\vskip 2mm

\noindent Recall that we always determine $LCR$ and $K$-rule compliancy after any intersection. Thus, the intersection: $I_{x_{1}, c_{2}} \cap I_{b_{3}, -c_{4}}$ = $0$, because at least one cell violated the $K$-rule. Here, both $C_{2}$ and $C_{4}$  violated the $K$-rule.\\

\vskip 2mm

\noindent The $LCR$ and $K$-rule compliant union: $I_{x_{1}, c_{2}} \cup I_{b_{3}, -c_{4}}$ is:

\vskip 2mm

\noindent   \hspace{1.1cm}  $( 0_{a}, 1_{x}, 1_{y} \hspace{.15cm}\textcolor{brown}{|} \hspace{.15cm}1_{-a}, 1_{b}, 1_{c} \hspace{.15cm}\textcolor{brown}{|} \hspace{.15cm}  1_{-a}, 1_{b}, 0_{-c} \hspace{.15cm}\textcolor{brown}{|} \hspace{.15cm} 1_{-a}, 1_{-b}, 1_{-c}    \hspace{.15cm}\textcolor{brown}{|} \hspace{.15cm}   1_{-a}, 1_{-b}, 1_{c} )$\\

\noindent Observe that $I_{x_{1}, c_{2}} \cup I_{b_{3}, -c_{4}}$ is a multi-edged sequence.  We remark that if a collection of edge-sequences are  $LCR$ and $K$-rule compliant, then their union must also be  $LCR$ and $K$-rule compliant, so compliancy is assured after a union is performed. The reason is simply that there is no union of $LCR$ and $K$-rule compliant edge-sequences, that could create a new $LCR$ scenario,  or cause a violation of the $K$-rule, via the Boolean rules.

\section{Description of the algorithm}

\vskip 2mm
\noindent In this section we describe the basic algorithm.  The description of the algorithm will consist of describing pre-processing and how the ordered sequences are compared with each other and what actions are to be taken based on those comparisons. The section to follow proves the correctness of this scheme  and that it stops in polynomial time, when applied to any instance of $3$-SAT.\\

\vskip 2mm

\begin{center}
\large { \textbf{Pre-processing}}

\end{center}

\noindent Pre-processing begins by building the first tables from a given DIMACS file. While doing  this, we learn of redundancies that might as well be eliminated. However, it's worth noting that the algorithm to process a $3$-SAT does not require any redundancies to be removed.\\

\begin{definition}\label{definition:  pure literal}
Let $\mathcal{T}$ be a  given collection of clauses. We call a literal $x$, a pure literal, if its negation $-x$, does not appear in $\mathcal{T}$. ie. $\nexists$ $-x$  in $\mathcal{T}$.

\end{definition}
 \vskip 2mm
\begin{definition}\label{definition:  pure literal}
A clause of the form: $(l, -l, x)$, $(l, -l)$ or $(l, l, -l)$  is called a quantum clause. A quantum clause is a possible randomly generated clause.
\end{definition}

\noindent To start, we order the literals represented by integers,  within each clause.  If there exists among the given clauses, two or more clauses which are the same clause, but the literals appear in a different order, it will be discovered. Only one copy of each unique clause is required to process a given $3$-SAT. We learn if any clause has literal duplication: $(a, a, a)$ or $(a, b, b)$ or $(a, a)$. Randomly generated clauses could be of this form, which we reduce to: $(a)$, $(a, b)$ and $(a)$, respectively. When we have ordered the clauses, we have also documented each literal and its negation, and to which clauses they are associated. At this point we will have discovered any $\it{pure}$ literals. The clauses containing one or more $\it{pure}$ literals, can be removed, where one of the  $\it{pures}$, represents its clause. We can remove a clause with a $\it{pure}$ because any solution found with the remaining clauses, can include  all the $\it{pures}$, since no solution includes their negations. We will also discover any $\it{quantum}$ clauses, and they can be removed because any solution for the  remaining clauses can always add a literal from a $\it{quantum}$. For example, let a $\it{quantum}$ clause be $q$ $=$ $(l, -l, x)$, and there is a solution with the remaining clauses which may use $l$ or $-l$ or neither. For neither, $l$ or $-l$ can be selected to be part of the solution.\\

\noindent Removing $\it{pures}$ and $\it{quantums}$ can be recursive, as it was for our example.\\

\noindent If there exists among the collection of clauses given, one or more clauses containing just one literal, we remove these clauses. Obviously, the single literal is the only literal to represent the clause for any solution. And since these literals must be in every solution, their negations are removed from the collection of clauses given. This action may also be recursive. If this action empties a clause of all its literals, then there is no solution for the collection of clauses given. \\

\noindent There is a possible redundancy that we ignore, which we call $\it{Global}$ $\it{Inclusion}$. Suppose among the given clauses, there exist two clauses: $C_{i}$ $=$ $(a, b, x)$ and $C_{j}$ $=$ $(a, b)$. It's clear that if a solution exists, then either literal $a$ or $b$ must be part of that solution, thus $C_{i}$ is redundant and could be removed. With respect to $\it{Global}$ $\it{Inclusion}$, in context of processing a $3$-SAT, a cell  $C_{k}$ may no longer have $1$ entries for every literal of clause  $C_{k}$. For example, let clause $C_{k}$ have three literals $a$, $b$ and $z$. Now, suppose after some processing, that the entry for $z$ in cell $C_{k}$ is zero, in all edge-sequences. Then cell  $C_{i}$ can be removed from all sequences. And after all processing is complete, it can be determined if $a$ or $b$ will be representing  clause $C_{i}$. \\

\begin{center}
\large { \textbf{Construction of the edge and vertex sequences}}

\end{center}

\noindent  Let $n$ $\le$ $3c$, where $c$ is the number of remaining clauses and $n$ is the sum of the sizes, for the $c$ clauses. Then after pre-processing, the $n$ vertex-sequences, grouped by clause association, are  constructed first, as outlined in definition $2.6$, and then $LCR$ and $K$-rule is applied. It would be common practice to order the clauses, and the literals within each clause, and use this same ordering for both the edge and vertex sequences. \\

\noindent After pre-processing, the remaining clauses not removed, have at most, $9$ edge-sequences constructed from them pairwise. The edge-sequences are constructed as outlined in definition $2.2$, and then $LCR$ and $K$-rule is applied. Observe that the number of edge-sequences would be less than ${n}\choose{2}$. It must always be less than, because we did not subtract the over count of non-existent edges with i)\hspace{.05cm} both endpoints in the same clause or ii)\hspace{.05cm} the non edge-sequences between a literal and its negation. Of course, if each clause had just one literal and there was a solution, pre-processing would have presented the solution or pre-processing removed all literals from at least one clause, establishing unsatisfiability. Either way, no edge-sequences would have been constructed. Each pair of clauses from the collection of $c$ clauses, forms an $S$-set. Thus, the number of $S$-sets is ${c}\choose{2}$, where any $S$-set contains at most, $9$ edge-sequences. We shall denote an $S$-set with the indices of the two clauses used to construct its edge-sequences. ie. $S_{i, j}$ has edge-sequences whose endpoints are from clauses $C_{i}$ and $C_{j}$.  \\

\noindent Before we describe in detail the \textbf{Comparing} of $S$-sets, we list the refinement rules.  The refinement rules below, refer to a bit-change occurring, an edge removed from its $S$-set, or a literal having a zero entry in every edge and vertex sequence, all being  the result of the Comparing of $S$-sets. Given a collection of clauses, if any of these refinements occurred that were not the result of the Comparing of $S$-sets, it was the result of $LCR$ and $K$-rule compliancy, while constructing the edge-sequences and vertex-sequences. We shall apply the actions outlined in the refinement rules even when constructing the vertex and edge sequences. The example demonstrated that certain edge-sequences were zero upon construction, as they failed $LCR$ and $K$-rule compliancy. To be systematic, we would first construct all the vertex-sequences as described in definition $2.6$. Then apply $LCR$ and $K$-rule to all of the vertex-sequences. If the actions outlined in the refinement rules can be taken on any vertex-sequence, we do so. Next, we  construct all the edge-sequences as described in definition $2.2$. Then apply $LCR$ and $K$-rule to all of the edge-sequences. If the actions outlined in the refinement rules can be taken on any vertex or edge sequence, we do so. This process may be recursive.\\

\noindent It is to be understood that an edge-sequence $I_{x, y}$ may be called just an edge for simplicity and that a vertex may be referred to by its associated literal. Also, an endpoint for an edge may be referred to as an endpoint for its corresponding edge-sequence and to say: a literal $x$ or an endpoint, in a cell $k$, is to be understood as a $1$ entry in cell $k$, associated to literal $x$ or an endpoint. \\

\newpage

\begin{center}
\large { \textbf{ The Refinement Rules $1, 2, 3$ and $4$}}

\end{center}

\noindent When a refinement occurs and one of the rules $1, 2, 3$ or $4$ actions are to be applied to an edge-sequence or a vertex-sequence,   $LCR$ and $K$-rule compliancy must always be determined afterward. Note that the actions for the refinement rules could be recursive and that all refinements are permanent.\\

\begin{center}
\textbf{Rule 1: a bit-change occurred }

\end{center}

\noindent Let  $I_{a_{i},b_{j}}$ where $i$ denotes cell $i$, and $j$ denotes cell $j$, be an edge-sequence  where at least, one $1$ entry for a position (occurrence) associated with literal $c$ became zero.
Then we change $1$ entries to zero for all positions associated with $c$ in $I_{a_{i},b_{j}}$. If there was just one position for $c$, then no action is taken.\\

\noindent Now, we consider all edges $I_{a, b}$ that are not $I_{a_{i},b_{j}}$. ie. at least one endpoint is not from either cell $i$ or $j$. We will change a $1$ to zero for every occurrence of a literal who had at least one occurrence become zero in $I_{a_{i},b_{j}}$  for all edge-sequences labelled $I_{a, b}$. The result is all  $I_{a, b}$ which of course includes $I_{a_{i},b_{j}}$  have the same set of literals whose associated entries are zero. \\

\begin{center}
\textbf{Rule 2: the follow up to a bit-change }

\end{center}

\noindent If  $I_{a,b}$ has zero entries for $c$, then  $I_{a,c}$ will have zero entries for $b$, and  $I_{b,c}$ will have zero entries for $a$.
All three edge-sequences are documenting the same fact, namely, that there is no solution with $a$, $b$ and $c$ together.
So, if  an edge-sequence $I_{a,b}$ for which at least one entry became zero, say for literal $c$, we apply rule $1$. Then in  $I_{b,c}$ and  $I_{a,c}$   we make all $1$ entries zero, that correspond to literals $a$ and $b$, respectively.\\

\begin{center}
\textbf{Rule 3: $I_{x_{i},y_{j}}$ $=$ $0$, and was removed from its $S$-set $S_{i, j}$ }

\end{center}

\noindent Let $I_{x_{i},y_{j}}$ be an edge that was removed from $S$-set, $S_{i, j}$. The sub-subscripts for the edge-sequence  $I_{x_{i},y_{j}}$  indicate that one endpoint is from cell $i$, and the other endpoint is from cell $j$.
First, we remove all occurrences of $I_{x_{m},y_{n}}$  where  $m$$+$$n$ $\ne$  $i$$+$$j$, from their respective $S$-sets.\\

\noindent  Next, we update all the vertex-sequences for literals $x$ and $y$ by:
All the vertex-sequences for $x$ have all positions associated with $y$ incur a bit-change.
And all the  vertex-sequences for $y$ have all positions associated with $x$ incur a bit-change.\\

\noindent Recall, that any kind of refinement requires testing $LCR$ and $K$-rule compliancy, even on vertex-sequences. This in turn could cause the vertex-sequence to equal zero. In such a case, we apply all actions outlined in rule $4$. \\

\noindent We also ensure that all refined vertex-sequences for a literal $x$, have the same set of literals whose associated entries are zero. \\

\noindent Now,  all positions associated with $y$ incur a bit-change, in $I_{x,\#}$  where  {\footnotesize$\#$} is any literal that forms an edge with $x$.
Then all positions associated with $x$ incur a bit-change, in $I_{\%,y}$  where {\footnotesize$\%$} is any literal that forms an edge with $y$.\\

 \noindent This whole process may cause another edge removal, in which case we apply rule $3$, recursively if need be.  This in turn could cause a vertex-sequence to equal zero. In such a case, we apply all actions outlined in rule $4$.\\

\begin{center}
\textbf{Rule 4: a literal's vertex-sequence equals zero }

\end{center}

\noindent Let a literal $x$ be such that its vertex-sequence violates the $K$-rule. Then its vertex-sequence $V_{x_{i}}$, where $x$ belongs to cell $C_{i}$ is equal to zero. \\

\noindent If $V_{x_{i}}$ is now zero, then all $V_{x}$ vertex-sequences are now evaluated to be zero, and all of them are removed from the vertex-sequence table.\\

\noindent When a vertex-sequence equals zero we do:  All positions associated with  $x$ incurs a bit-change in every vertex and edge sequence. This action in turn causes all edge-sequences of the form: $I_{x,\#}$, where {\footnotesize$\#$} is any literal that forms an edge with $x$, to equal zero.  \\

\noindent  This whole process may cause another edge-sequence (not of the form $I_{x,\#}$), to be removed, in which case we apply rule $3$, recursively if need be.  This in turn could cause a vertex-sequence to equal zero. In such a case, we apply all actions outlined here, in rule $4$.\\

\newpage

\begin{center}
\Large {\textbf{The Comparing of the $S$-sets}}

\end{center}

\vskip 2mm

\noindent Essentially, the Comparing process is the algorithm.
All data structures are simply updated based on the outcome of Comparing $S$-sets with one another. \\

\begin{definition}\label{definition:  a run}
When every $S$-set has been Compared with every other $S$-set, we say that a \textbf{run} has been completed. If $c$ clauses are considered, then there are  ${c}\choose{2}$\hspace{.1cm} $S$-sets, thus the number of $S$-set comparisons for a \textbf{run} is ${{{c}\choose{2}}\choose{2}}$ $<$ $c^{4}$.
\end{definition}

\begin{definition}\label{definition:  a round}

A \textbf{round} is completed if the Comparing  process stops because no refinement occurred during an entire \textbf{run}.  We say that the $S$-sets are \textbf{equivalent} when a round is completed.
\end{definition}

\noindent We note that if the first round was not completed, it was the case that the vertex-sequences associated to  a clause, were evaluated to be zero, so they were discarded. This violation of the $K$-rule stops all processing, as there is no solution for the collection of clauses given.\\

\noindent To Compare, we take two $S$-sets and \textbf{determine} if an edge-sequence $I_{x, y}$ from one of the $S$-sets, can  be refined by a union of the intersections between $I_{x, y}$  with each of the edge-sequences, from the other $S$-set. Either $I_{x, y}$  the edge-sequence under determination, is refined or it remains the same. This is done for each edge-sequence from both of the $S$-sets, in the same manner. As a matter of practice, we determine in turn, each edge-sequence from one $S$-set first, and then  we determine in turn, each edge-sequence from the other $S$-set. Below, we construct two $S$-sets to describe in more detail all the steps to be taken.\\

\noindent Let the $S$-set:  $S_{i, j}$ \hspace{.03cm}contain $9$ edge-sequences with endpoints from clauses: $C_{i}$ $=$ $(1, 2, 3)$ and $C_{j}$ $=$ $(a, b, c)$\hspace{.03cm} giving: $I_{1, a}$, $I_{1, b}$, $I_{1, c}$, $I_{2, a}$, $I_{2, b}$, $I_{2, c}$, $I_{3, a}$, $I_{3, b}$, $I_{3, c}$ \\

\noindent Let the $S$-set: $S_{k, l}$ \hspace{.03cm}contain $9$ edge-sequences with endpoints from clauses: $C_{k}$ $=$ $(4, 5, 6)$ and $C_{l}$ $=$ $(d, e, f)$\hspace{.03cm} giving: $I_{4, d}$, $I_{4, e}$, $I_{4, f}$, $I_{5, d}$, $I_{5, e}$, $I_{5, f}$, $I_{6, d}$, $I_{6, e}$, $I_{6, f}$ \\

\noindent If $S_{i, j}$ and $S_{k, l}$ have $9$ edge-sequences each, then there were no negations between the literals in $C_{i}$ and $C_{j}$, or
between the literals in $C_{k}$ and $C_{l}$, respectively.\\

\noindent We say that \textbf{determining} all edge-sequences from one $S$-set first, is doing one direction denoted by: $S_{k,l}$ $\overset{1}{\rightharpoonup}$ $S_{i,j}$. And determining all edge-sequences once, for both $S$-sets,  is doing both directions, denoted by:  $S_{k,l}$ $\overset{1}{\underset{2}{\rightleftharpoons}}$ $S_{i,j}$ \\

\noindent Suppose we \textbf{determine} $I_{4, d}$ of  $S_{k, l}$ first. Then we have: \\

\noindent $(I_{1, a}\cap \textcolor{blue}{I_{4, d}})\cup (I_{1, b}\cap \textcolor{blue}{I_{4, d}})\cup (I_{1, c}\cap \textcolor{blue}{I_{4, d}}) \cup (I_{2, a}\cap \textcolor{blue}{I_{4, d}}) \cup  (I_{2, b}\cap \textcolor{blue}{I_{4, d}})\hspace{.05cm} \cup $ \\
\noindent $(I_{2, c}\cap  \textcolor{blue}{I_{4, d}})  \cup  (I_{3, a}\cap \textcolor{blue}{I_{4, d}}) \cup (I_{3, b}\cap \textcolor{blue}{I_{4, d}}) \cup (I_{3, c}\cap \textcolor{blue}{I_{4, d}}) \hspace{.1cm} \le   \hspace{.1cm} \textcolor{blue}{I_{4, d}} $ \\

\noindent \textbf{The  $\Phi$ rule}: Wrt. an intersection between  $I_{4, d}$ and an edge-sequence from $S_{i, j}$, suppose the intersection resulted with occurrences of literal $z$ and its negation $-z$, having zero entries, where at least one occurrence is not in any of the endpoint cells. Then the intersection can be evaluated to be zero, since if these two edge-sequences belonged to a solution together, they would also belong to two other solutions, one using $z$ and another using $-z$, if both $z$ and $-z$ had occurrences not in the endpoint cells. \\

\noindent The one efficiency present even in the original $\it{naive}$ version with no refinement rules, was eliminating any unnecessary intersections by one easy check. After an edge-sequence is selected for determination, say $I_{x_{r}, y_{s}}$ where $x_{r}$ $\in$ $C_{r}$, $y_{s}$ $\in$ $C_{s}$, the edge-sequences from the other $S$-set in the Comparing, that do not have $1$ entries for the endpoints $x_{r}$ and $ y_{s}$  when intersected with $I_{x_{r}, y_{s}}$, will be zero. Recall, that the two cells containing the endpoints, only have a single $1$ entry corresponding to the two endpoints' positions, in their respective cells. Thus, if the other edge-sequence does not have a $1$ entry for those same positions, the intersection will be zero, due to $K$-rule violation. So, after the selection of an edge-sequence to be determined, we select the edge-sequences from the other $S$-set, if they have $1$ entries in both positions corresponding to the endpoints of $I_{x_{r}, y_{s}}$. Of course, we could also check to see if there are $1$ entries in $I_{x_{r}, y_{s}}$ corresponding to the endpoints of the other edge-sequence as well. \\

\noindent Let's suppose then, that every edge-sequence in $S_{i, j}$ above, did have $1$ entries for both endpoints  $4_{k}$ and $ d_{l}$ of $I_{4, d}$. Now suppose $6$ intersections become zero, after the intersections were taken and $LCR$ and $K$-rule was applied to each intersection, and we now have: \\

\noindent $ 0 \cup (I_{1, b}\cap \textcolor{blue}{I_{4, d}})\cup 0 \cup (I_{2, a}\cap \textcolor{blue}{I_{4, d}}) \cup  0 \cup 0  \cup  0 \cup 0 \cup (I_{3, c}\cap \textcolor{blue}{I_{4, d}}) \hspace{.1cm} \le   \hspace{.1cm} \textcolor{blue}{I_{4, d}} $ \\

\noindent which is equivalent to:  $  (I_{1, b}\cap \textcolor{blue}{I_{4, d}})\cup  (I_{2, a}\cap \textcolor{blue}{I_{4, d}})  \cup (I_{3, c}\cap \textcolor{blue}{I_{4, d}}) \hspace{.1cm} \le   \hspace{.1cm} \textcolor{blue}{I_{4, d}} $ \\

\noindent Now, we take their union. Of course, $LCR$ and $K$-rule compliancy need not be checked for any union, since it would not have been possible to create a new loner cell scenario, nor a cell with all zero entries.\\

\noindent Then to complete the determination of $I_{4, d}$ we need to compare position by position, to see if $I_{4, d}$ has been refined. More precisely, if we have: $  (I_{1, b}\cap \textcolor{blue}{I_{4, d}})\cup  (I_{2, a}\cap \textcolor{blue}{I_{4, d}})  \cup (I_{3, c}\cap \textcolor{blue}{I_{4, d}}) $ $ = $  $ \textcolor{blue}{I_{4, d}} $, then $I_{4, d}$ is unchanged, and we move on to the  next edge-sequence to be determined. Or instead, we have: $  (I_{1, b}\cap \textcolor{blue}{I_{4, d}})\cup  (I_{2, a}\cap \textcolor{blue}{I_{4, d}})  \cup (I_{3, c}\cap \textcolor{blue}{I_{4, d}}) $ $ < $  $ \textcolor{blue}{I_{4, d}} $, then $I_{4, d}$ has been refined. We need to know which literals incurred a bit-change, and then we apply the appropriate actions of the refinement rules, and as always, followed by testing $LCR$ and $K$-rule compliancy.\\

\noindent To summarize, an edge-sequence $I_{4, d}$  to be \textbf{determined}, has $4$ possible scenarios.\\

\noindent 1) $I_{4, d}$ $\ne$ $0$, and is unchanged.\\

\noindent 2)  $I_{4, d}$ $\ne$ $0$, and is refined. We determine which literals had a bit-change and follow all appropriate refinement rule actions, recursively if need be.\\

\noindent 3)  $I_{4, d}$ $=$ $0$, because $  (I_{1, b}\cap \textcolor{blue}{I_{4, d}})\cup  (I_{2, a}\cap \textcolor{blue}{I_{4, d}})  \cup (I_{3, c}\cap \textcolor{blue}{I_{4, d}}) \hspace{.1cm} \le   \hspace{.1cm}   \textcolor{blue}{I_{4, d}} $ became zero after  an appropriate refinement rule action was taken, and then $LCR$ and $K$-rule was applied. In this case, edge-sequence $I_{4, d}$ is discarded, and the actions stated in refinement rule $3$ are taken, recursively if need be. \\

\noindent 4) If  each intersection: $  (I_{1, b}\cap \textcolor{blue}{I_{4, d}})$,  $(I_{2, a}\cap \textcolor{blue}{I_{4, d}})$ and  $(I_{3, c}\cap \textcolor{blue}{I_{4, d}})$ had also been zero at the outset, then $I_{4, d}$ $=$ $0$. And as with 3), $I_{4, d}$ is discarded and the actions stated in refinement rule $3$ are taken, recursively if need be. When an edge-sequence equals zero, it was the case that none of the edge-sequences from  an $S$-set could be part of a solution with the edge-sequence that was being determined. And every solution has an edge-sequence in every $S$-set. \\

\newpage

\noindent There are two observations worth noting with respect to the Comparing process. First, that the result of an intersection with just the edge-sequence to be determined, or the union of  the intersections between the edge-sequence to be determined, with each of the allowed edge-sequences, from the other $S$-set, is never a multi-edged sequence. Whether no refinement took place for the edge-sequence being determined, or a great deal of refinement took place, it is still the edge-sequence  that is being determined and it does indeed represent an edge-sequence with two particular endpoints. \\

\noindent In section $2$,  we gave an example of an intersection and a union (from Figure $1$), but neither was for an edge-sequence determination. Recall, that the intersection violated the $K$-rule, thus equals zero, but the union was not evaluated to be zero, and it's properly defined as a multi-edged sequence. \\

\noindent The second observation is that the distributive law for unions and intersections does not apply here. Otherwise, for example,  we could say that: $  (I_{1, b}\cap \textcolor{blue}{I_{4, d}})\cup  (I_{2, a}\cap \textcolor{blue}{I_{4, d}})  \cup (I_{3, c}\cap \textcolor{blue}{I_{4, d}}) \hspace{.1cm} =   \hspace{.1cm}   (I_{1, b} \cup  I_{2, a}  \cup I_{3, c}) \cap \textcolor{blue}{I_{4, d}}$ which is not true in general. In the algorithm, we provide a Comparing between two $S$-sets that will do more than once in each direction. If doing the second direction, $S_{k,l}$ $\overset{2}{\leftharpoondown}$ $S_{i,j}$ results in one or more bit-changes, we will do  $S_{k,l}$ $\overset{3}{\rightharpoonup}$ $S_{i,j}$ again. And if one or more  bit-changes occur in $S_{k,l}$ $\overset{3}{\rightharpoonup}$ $S_{i,j}$  we will do  $S_{k,l}$ $\overset{4}{\leftharpoondown}$ $S_{i,j}$ again, and so on until no bit-change occurs while doing a direction. An important note, is that we don't have to do this. If we only did $S_{k,l}$ $\overset{1}{\underset{2}{\rightleftharpoons}}$ $S_{i,j}$  no matter how many bit-changes occurred while doing the second direction, it wouldn't matter.  Suppose doing  $S_{k,l}$ $\overset{3}{\rightharpoonup}$ $S_{i,j}$ would have had a bit-change. Then if that scenario still exists in the next \textbf{run}, meaning that some other Comparing of $S$-sets didn't do the refinement in question prior to returning to these two $S$-sets Comparing, a bit-change would still occur for $S_{k,l}$ ${\rightharpoonup}$ $S_{i,j}$. We are assuming by doing more than just  $S_{k,l}$ $\overset{1}{\underset{2}{\rightleftharpoons}}$ $S_{i,j}$ if prompted by a bit-change occurring during $S_{k,l}$ $\overset{2}{\leftharpoondown}$ $S_{i,j}$ is an efficiency measure. Recall definition $3.3$, that after all pairs of $S$-sets have been compared with each other, a \textbf{run} has been completed, which is ${{{c}\choose{2}}\choose{2}}$ $<$ $c^{4}$, for $c$ clauses. Another \textbf{run} will commence if any refinement occurred. Eventually, a \textbf{run} will incur no refinement, not even a bit-change, at which point a \textbf{round} has been completed, and the $S$-sets are said to be \textbf{equivalent}, which implies satisfiability. Or, the algorithm stopped because unsatisfiability had been discovered during  \textbf{round} $1$. Discovering unsatisfiability is when every literal from some clause is such that their vertex-sequences equal zero.

\begin{center}
\large {\textbf{An example of Comparing}}

\end{center}

\noindent The two \textbf{Comparings} shown in this example, produce the same outcome as the action for refinement rule $2$. \\

\noindent  Let the $S$-set:  $S_{i, j}$ \hspace{.03cm}contain $9$ edge-sequences with endpoints from clauses: $C_{i}$ $=$ $(a, 2, 3)$ and $C_{j}$ $=$ $(b, 4, 5)$\hspace{.03cm} giving: $I_{a_{i}, b_{j}}$, $I_{a, 4}$, $I_{a, 5}$, $I_{2, b}$, $I_{2, 4}$, $I_{2, 5}$, $I_{3, b}$, $I_{3, 4}$, $I_{3, 5}$ \\

\noindent Now suppose we have clause: $C_{k}$ $=$ $(c, 6, 7)$. Then $I_{a_{i}, c_{k}}$ $\in$  $S_{i, k}$ and  $I_{b_{j}, c_{k}}$ $\in$  $S_{j, k}$ Further suppose that $I_{a_{i}, b_{j}}$ incurred a bit-change for literal $c$. \\

\noindent First we consider:  $S_{i,j}$ $\overset{2}{\underset{1}{\rightleftharpoons}}$ $S_{i,k}$  where we determine $I_{a_{i}, c_{k}}$\\

\noindent Observe that all edge-sequences: $I_{2, b}$, $I_{2, 4}$, $I_{2, 5}$, $I_{3, b}$, $I_{3, 4}$, $I_{3, 5}$ have a zero entry for $a_{i}$, so they are not selected for the determination of $I_{a_{i}, c_{k}}$, again because their intersection would be just zero. Since,  $I_{a_{i}, b_{j}}$ has zero entries for all occurrences of $c$, then $I_{a_{i}, b_{j}} \cap I_{a_{i}, c_{k}}$ $=$ $0$. Thus, the only two edge-sequences from $S_{i,j}$ to determine $I_{a_{i}, c_{k}}$ is:  $I_{a, 4}$ and $I_{a, 5}$, both of which have a zero entry for $b_{j}$. Therefore, the determination of $I_{a_{i}, c_{k}}$ is a refinement where at least $b_{j}$ has a zero entry. Then after refinement rule $1$ is applied, all occurrences of literal $b$ in $I_{a_{i}, c_{k}}$ are zero, assuming $I_{a_{i}, c_{k}}$ $\ne$ $0$, after $LCR$ and $K$-rule compliancy.  \\

\noindent Similarly, we consider:  $S_{i,j}$ $\overset{2}{\underset{1}{\rightleftharpoons}}$ $S_{j,k}$  where we determine $I_{b_{j}, c_{k}}$\\

\noindent Observe that all edge-sequences:  $I_{a, 4}$, $I_{a, 5}$,  $I_{2, 4}$, $I_{2, 5}$,  $I_{3, 4}$, $I_{3, 5}$  have a zero entry for $b_{j}$, so they are not selected for the determination of $I_{b_{j}, c_{k}}$.  Since,  $I_{a_{i}, b_{j}}$ has zero entries for all occurrences of $c$, then $I_{a_{i}, b_{j}} \cap I_{b_{j}, c_{k}}$ $=$ $0$. Thus, the only two edge-sequences from $S_{i,j}$ to determine $I_{b_{j}, c_{k}}$ is:  $I_{2, b}$ and $I_{3, b}$, both of which have a zero entry for $a_{i}$. Therefore, the determination of $I_{b_{j}, c_{k}}$ is a refinement where at least $a_{i}$ has a zero entry. Then after refinement rule $1$ is applied, all occurrences of literal $a$ in $I_{b_{j}, c_{k}}$ are zero, assuming $I_{b_{j}, c_{k}}$ $\ne$ $0$, after $LCR$ and $K$-rule compliancy.  \\

\noindent \textbf{Summary}: Pre-processing begins with a DIMACS file  submission, providing clauses, where each clause is at most size $3$. The pre-processing entails ordering the literals  within each clause (by any convention desired), and then removing duplicate clauses or literals within a clause.  Next, the clauses to be removed  have at least one $\it{pure}$ literal, or they are quantum clauses. This may be a recursive process. Finally, clauses of size $1$, a $\it{loner}$ clause, are removed as are the negations of the literals from these clauses of size $1$ from all the other clauses that were given. This may also be recursive, as a negation may have been in a clause of size $2$, thus its removal created a new loner clause. At this point, edge and vertex sequences are constructed from the remaining clauses. The edge-sequences are grouped in their respective $S$-sets and the vertex-sequences are grouped by their clause association in the vertex-sequence table. When the vertex and edge sequences are $LCR$ and $K$-rule compliant, the Comparing process begins. The Comparing process stops if: i)  a clause was such that all its literals' vertex-sequences are zero, where it's reported that the given collection of clauses has no solution. Or, ii) one or more \textbf{runs} take place, where the last \textbf{run} had no refinement. This signals the end of a round, and the $S$-sets are said to be  \textbf{equivalent}, which implies satisfiability. \\

\noindent \textbf{Claim }:  If a $3$-SAT  $\mathcal{G}$, with $c$ clauses (after pre-processing), has a solution $K_{C}$, then the $S$-sets for $\mathcal{G}$, will be equivalent at the completion of round $1$, where each edge-sequence $I_{x,y}$ that appears, belongs to at least one solution.  Moreover,  an edge-sequence $I_{x, y}$ is such that any literal with a $1$ entry in $I_{x, y}$ belongs to at least one $K_{C}$ with $x$ and $y$. Lastly, if at least one clause is such that all its literals' vertex-sequences are zero, it means there is no solution for the collection of clauses that were processed, thus unsatisfiability has been discovered, which stops the processing of round $1$. \\

\noindent In the next section we establish the claim.  This is followed by proving that attaining $S$-set equivalency is always achieved in polynomial time. \\

\section{Proof of correctness and termination}

\vskip 4mm

\begin{center}
 \textbf{Comparing and the Refinement rules}

\end{center}

\noindent The actions of the refinement rules are taken if one of the four refinements occurred,  by the Comparing process.  We assert that:  1) If by Comparing, a bit-change occurred for one position of a literal $x$ in some edge-sequence $I$, the Comparing process will eventually cause a bit-change for every position of $x$ in $I$. 2)  All edge-sequences $I_{x, y}$ have the same set of literals with zero entries via Comparing. 3) If by Comparing, an edge-sequence is evaluated to be zero, then all edge-sequences whose endpoints have the same associated literals, will be evaluated to be zero, by Comparing. And finally, 4) If by Comparing, a literal's vertex-sequence $V_{x_{i}}$ is evaluated to be zero, then  all vertex-sequences for literal $x$ belonging to other clauses, will be evaluated to be zero, by Comparing. Note that we have already shown in the Comparing example, that refinement rule $2$ is done by Comparing.\\

\noindent Since we intend to prove the algorithm that does use the refinement rules, then to determine if its actions are merely an efficiency isn't warranted. Moreover, the argument to use the actions of the refinement rules ensures contradiction free edge-sequences. More precisely, if a bit-change occurred for one  occurrence of literal $c$ in edge-sequence $I_{a, b}$ by Comparing, then with respect to literals, there is no solution using literals $a$, $b$ and $c$ together. Therefore, if another  occurrence of literal $c$ exists in $I_{a, b}$ it's still the case that there is no solution using literals $a$, $b$ and $c$ together. Thus, there would be no purpose for other occurrences of $c$ in  $I_{a, b}$  to have $1$ entries. It follows, that if there exists another $I_{a, b}$ that belongs to another $S$-set, that  there would be no purpose for any occurrence of $c$  to have $1$ entries in this $I_{a, b}$. And if by Comparing, $I_{a, b}$ $=$ $0$, then with respect to  literals, there is no solution using literals $a$ and $b$ together, so all edge-sequences whose endpoints are associated with literals $a$ and $b$ should be zero to be contradiction free. And finally, if a literal's vertex-sequence  $V_{x_{i}}$ $=$ $0$, by Comparing, meaning no solution exists using literal $x$, then with respect to  literals, the vertex-sequence for every occurrence of $x$ should be zero to be contradiction free.  We would provide the proof of the assertion if there should be any interest. \\

\noindent Another efficiency rule of note, but not part of our algorithm, is with respect to the vertex-sequences. Suppose during the Comparing process, it is determined that $V_{x}$ $=$ $0$ say, and then at some later time  in the Comparing process,  it is determined that $V_{-x}$ $=$ $0$. If $\mathcal{G}$ were to be satisfiable, then as explained in the $\Phi$ rule, a contradiction would exist. Therefore, $\mathcal{G}$ must be unsatisfiable, so the processing can stop.

\begin{lemma}\label{lemma: no edge-sequence belonging to a $K_{C}$ is removed.}
 Let  $\mathcal{W}$ be a collection of $S$-sets with $LCR$ and $K$-rule compliant edge-sequences of length $c$ cells. Then no edge-sequence from  $\mathcal{W}$, belonging to a $K_{C}$, is removed by the Comparing process.

\end{lemma}

\proof
Suppose there exists a $K_{C}$, $\mathcal{M}$. Then there exists an edge-sequence in each $S$-set of $\mathcal{W}$, such that \hspace{.05cm}$\underset{i, j}{\bigcap}\hspace{.05cm} I_{i, j}$ $\ne$ $0$, where $i$ and $j$ are every pair of endpoints for the collection of edge-sequences of  $\mathcal{M}$, by definition $2.12$. So, if we take any edge-sequence $I$, belonging to $\mathcal{M}$, it will still exist after the determination of $I$ from every other $S$-set. The reason is that each $S$-set $S_{i, j}$, has an edge-sequence $I_{x_{i},  y_{j}}$, call it $I'$, of  $\mathcal{M}$, and the intersection $I \cap I'$, preserves all the $1$ entries for every literal that is part of the solution, by definition of $\mathcal{M}$. Thus, $I \cap I'$ will not equal zero after $LCR$ and $K$-rule compliancy, and the union of any additional intersections with $I$, for any given determination, will be of no consequence. Additionally, $I$ may be refined due to refinement rules taken on it or on other edge-sequences, but this will not effect the $1$ entries for the literals that belong to $\mathcal{M}$. That is, if any $1$ entry in $I$ for a literal of $\mathcal{M}$, did incur a bit-change, it would imply that another edge or endpoint of $\mathcal{M}$, equals zero, a contradiction. In other words, every determination of $I$ with all the  ${c}\choose{2} $$-$ $1$ $S$-sets will at most refine $I$,  since there is an edge-sequence belonging to $\mathcal{M}$, in every $S$-set, and their intersection, $I \cap I'$, preserves the $1$ entries for every literal that is part of the solution  $\mathcal{M}$. And no $1$ entry in $I$ for a literal of $\mathcal{M}$, is removed via the actions of the refinement rules on any edge-sequence, otherwise  $\mathcal{M}$ did not exist, a contradiction.   \\

\endproof

\noindent Observe that the proof for lemma $4.1$ establishes that if an edge-sequence $I_{x,y}$ and some literal $z$  belong to a $K_{C}$, then $z$ does not incur a bit-change in $I_{x,y}$  by Comparing. \\

\begin{lemma}\label{lemma: all singletons.}

Let  $\mathcal{X}$ be a collection of equivalent $S$-sets with edge-sequences of length $c$ cells, for a $3$-SAT $\mathcal{G}$ with $c$ clauses. Suppose an edge-sequence $I_{x,y}$ from $\mathcal{X}$,  is such that its $1$ entries correspond to just edge-singletons or singletons, which includes $x$ and $y$. Then $I_{x,y}$ belongs to at least one $K_{C}$.

\end{lemma}

\proof Let an edge-sequence be $I_{x,y}$. Observe, that if there is a collection $\mathcal{S}$ of $1$ entries, one from each cell, whose associated literals are such that no literal and its negation appear, then there is a solution by definition $2.1$. And by lemma $4.1$, the edges between those literals which did exist at the outset, will not be removed by the Comparing process, thus there is a corresponding $K_{C}$. ie. a solution as defined in $2.12$. \\

\noindent Let an edge-sequence $I_{x,y}$ with $c$ cells, that is $LCR$ and $K$-rule compliant, be such that its $1$ entries correspond to just edge-singletons or singletons.\\

\noindent We shall construct a collection $\mathcal{S}$  by first selecting the literals from all loner cells. This sub-collection is non-empty since endpoints $x$ and $y$ are in loner cells. Recall, that if $I_{x,y}$ is $LCR$ compliant, then the negations of the literals in loner cells do not exist in $I_{x,y}$. Assume that the collection $\mathcal{S}$ being constructed is not yet size $c$. Note that at this point, there are only cells with at least two $1$ entries, from which to select a literal. For selecting literals from such cells, we  make two cases to simplify the description of constructing a collection $\mathcal{S}$.\\

 \noindent \textbf{Case $i)$}: For edge-sequence $I_{x,y}$ there are no edge-pure literals in cells with two or three $1$ entries. \\

 \noindent \textbf{Step 1}: We select any cell $C_{i}$  and choose one of the literals, say $a$, to represent the cell. Next, we find the cell  that contains the $1$ entry for literal $-a$, say $C_{j}$. If cell $C_{j}$ has a $1$ entry for the negation of a literal in $C_{i}$, which is not representing $C_{i}$  then step $1$ stops. Suppose $C_{j}$ does not. Then there exist one or two $1$ entries for literals not $-a$, that are not the negations of any literal with a $1$ entry in $C_{i}$. Choose one of these literals to represent cell $C_{j}$ and find the cell $C_{k}$ that contains the $1$ entry for the negation of the literal chosen to represent $C_{j}$. If cell $C_{k}$ has a $1$ entry for the negation of a literal in $C_{i}$ or $C_{j}$  which is not representing $C_{i}$ or $C_{j}$ then step $1$ stops.
If this is not the case, then continue step $1$ as described above, until a cell is finally selected which does have a $1$ entry for a literal which is the negation of some literal in a previously selected cell, that does not represent that cell. Should this never occur, then the last cell selected is the $c^{th}$ cell, and there exists at least one $1$ entry for a literal that is not the negation of any literal that represents a cell.\\

\noindent \textbf{Step 2}: Suppose step $1$ stops and all $c$ cells have not been selected. Then the last cell selected has at least one $1$ entry for literal $z$ say, which is the negation of some literal in a previously selected cell, that does not represent that cell. The literal $z$ will represent the last cell selected. Now, we will find all cells not yet selected, that contain the negations of  literals in  previously selected cells which did not represent their cells. These negations will be chosen to represent the cells in which they were found. Of course, should more than one negation belong to the same cell, we  pick any one to represent the cell. Next, after selecting the cells that contain the negations of literals in previously selected cells, we check these cells to see if there are literals not representing their cells, that are not the negations of any literal among the cells selected thus far. If such literals exist, then we will find all cells not yet selected, that contain the negations of  these literals. These negations will be chosen to represent the cells in which they were found. We repeat this part of step $2$ as just described, until there are no literals of this kind. At this point, if all $c$ cells have not been selected, then a $\it{closed}$ $\it{circuit}$ exists, meaning the collection of selected cells (except the loner cells),  contain a literal and its negation. If this is the case, the collection of cells not yet selected, contain a literal and its negation as well. Therefore, we apply step $1$ again on this remaining collection of cells which have no literal representing them. Observe that no literal from the cells of the $\it{closed}$ $\it{circuit}$ are contained in these remaining cells. Step $1$ may stop again before all $c$ cells have been selected, in which case, step $2$ is applied. This may be recursive until each of the $c$ cells has a literal representing it, thus a $\mathcal{S}$ of size $c$ will have been constructed. \\

 \noindent \textbf{Case $ii)$}: For edge-sequence $I_{x,y}$  there exist, one or more cells having two or three $1$ entries, that contain at least one edge-pure literal.\\

 \noindent After selecting all literals from loner cells, we select all the edge-pure literals from  cells having two or three $1$ entries.
 These literals will represent their cell and if a cell has $1$ entries for more than one edge-pure literal, any one can be chosen to represent that cell. So, selecting cells with an edge-pure literal may leave the cells with no representative as yet, with edge-pure literals among them. Thus, this process may be recursive. If this recursive process is not all $c$ cells, then it is a $\it{closed}$ $\it{circuit}$, otherwise the process would continue. We are now in case $i)$ again, so we begin with step $1$ for the remaining cells which have no literal representing them thus far. Again, all of the above may be recursive until each of the $c$ cells has a literal representing it, producing a collection $\mathcal{S}$ of size $c$. Since we were able to construct a collection $\mathcal{S}$ of size $c$ for the two possible cases outlined above, it follows that $I_{x,y}$ belongs to at least one $K_{C}$.

\endproof

\noindent Note well that lemma $4.2$ implies that if the stated scenario occurs, in a $\mathcal{W}$ produced for a $3$-SAT $\mathcal{G}$, then the remaining processing  can finish with a linear time construction of a solution for $\mathcal{G}$. \\

\vskip 2mm

\begin{center}
 \textbf{  $S$-set collections and refinements }

\end{center}

\noindent We claim that no  $K_{C-1} \not\subset K_{C}$ for any collection of $S$-sets $\mathcal{W}$   with $LCR$ and $K$-rule compliant edge-sequences of length $c$ cells, constructed from the $c$ clauses for a $3$-SAT $\mathcal{G}$, exists. Suppose otherwise. Then  let  $\mathcal{M}$ be a $K_{C-1} \not\subset K_{C}$, and let clause $C_{i}$ $=$ $(a, b, c)$ be the clause that does not contain an endpoint for any edge of $\mathcal{M}$. Then at most, only two negations of the literals in $C_{i}$ could be used for $\mathcal{M}$, since all edge-sequences must be $LCR$ compliant. So, suppose $-a$ and $-b$ are endpoints for edges of  $\mathcal{M}$. Then there exists an edge-sequence  $I_{-a, -b}$ for $\mathcal{M}$. However,  $I_{-a, -b}$ has a zero entry for every occurrence of literal $-c$, otherwise, $I_{-a, -b}$ would not be $LCR$ compliant, thus, by definition $2.12$, $\mathcal{M}$ can't use literal $-c$. Since, there were edges between literal $c$ $\in$ $C_{i}$ with every literal who is an endpoint for all the edges of  $\mathcal{M}$, then by lemma $4.1$, the edges between those literals which did exist at the outset, will not be removed by Comparing. This implies that $\mathcal{M} \subset K_{C}$, contradicting the assumption that $\mathcal{M}$ was not contained. Observe, that the same argument  holds, no matter which two literals of $C_{i}$  had their negations as  endpoints for edges of $\mathcal{M}$. Also, if a  $K_{C-1}$  uses only one or no negation of a literal from the clause that does not contain an endpoint for any edge of $K_{C-1}$, then the $K_{C-1}$ is contained in some  $K_{C}$, by lemma $4.1$. \\

\noindent With respect to just literals, there is a scenario sometimes referred to as $\it{atomica}$, which is: given a collection of $m$ clauses, there is a solution for any choice of $m$$-1$ clauses, but no solution for all $m$ clauses. However, the $\it{atomica}$ scenario never occurs with any  $\mathcal{W}$, due to $LCR$ compliancy of the edge-sequences, as demonstrated above. ie. $\nexists$ a  $K_{C-1}$ that uses the negations of all the literals from the clause not containing an endpoint for the $K_{C-1}$. In other words, by construction, all  $K_{C-1} \not\subset K_{C}$ scenarios are  eliminated.\\

\noindent So, if a literal $x$ is such that it does not belong to any $K_{C}$ for some $\mathcal{W}$ with edge-sequences of length $c$ cells, then by the claim above, it's known that $x$ does not belong to any $K_{C-1}$ of $\mathcal{W}$, as well. Moreover, $x$ does not belong to a $K_{P}$, $p < c$$-1$, if there exist two or more clauses not having endpoints for any edge of the $K_{P}$, where all together, they have no literal and its negation among them. For example, suppose there are two clauses $C_{i}$ $=$ $(a, b, c)$ and \\$C_{j}$ $=$ $(d, e, f)$, where the literals from $C_{i}$ and $C_{j}$ together, do not have a literal and its negation.
Now, consider a $K_{P}$, so a $K_{C-2}$ denoted $\mathcal{L}$,  that has no endpoint for any of its edges in  $C_{i}$ or $C_{j}$. Suppose to the contrary that $x$ $\in$ $\mathcal{L}$. Now, $\mathcal{L}$ can at most, only use two negations for the literals of $C_{i}$, and at most, only two negations for the literals of $C_{j}$. Since, $C_{i}$ and $C_{j}$ together do not contain a literal and its negation, there exists a $K_{2}$  (an edge),  between  $C_{i}$ and $C_{j}$ for any possible choice of four negations of the literals from $C_{i}$ and $C_{j}$ which implies that $\mathcal{L} \subset  K_{C-1} \subset K_{C} \implies x \in K_{C}$, a contradiction.  \\

\noindent \textbf{Refinement Claim} : Suppose a collection of $S$-sets $\mathcal{W}$  incurred  one or more bit-changes called refinement $R$, by  Comparing, at some stage in the Comparing process. Now suppose we \textbf{imposed} in an arbitrary fashion, bit-changes to the edge-sequences of  $\mathcal{W}$, and call it refinement $S$,  prior to any Comparing. Then the same refinement $R$, would still occur by the Comparing of $\mathcal{W}$, or $R$ is fully or partly subsumed by $S$.  \\

\noindent Proof: We first note that refinements are due to the relationships between  the edge-sequences and conversely, refinements cause relationships between  the edge-sequences. So, if a collection of edge-sequences  cause a bit-change for a literal $c$, in  some edge-sequence $I_{a,b}$ during some order of Comparing, then imposing a bit-change to any of the edge-sequences involved, prior to any Comparing, will still result with no possibility of a $K_{C}$, using literals $a, b$ and $c$ together. More precisely, suppose an edge-sequence $I_{a,b}$ incurs a bit-change  for a literal $c$ by some order of Comparing. Now suppose that before any Comparing, we imposed any arbitrary refinement $S$. There are three cases to consider.\\

\noindent  The first case is that refinement $S$, played no role in the bit-change of $c$ in $I_{a,b}$. So, proceeding with the same order of Comparing, still results in a bit-change for $c$ in $I_{a,b}$. \\

\noindent The second case is when the refinement $S$, does play a role, where during the same sequence of Comparing, $S$ causes the bit-change of $c$ in $I_{a,b}$ to occur sooner, during the same order of Comparing. The refinement $S$ may have also caused other refinements, including to $I_{a,b}$ during  the same order of Comparing. \\

\noindent The third case is when the refinement $S$, does play a role by subsuming some or all of $R$. The most extreme would be that $S$ causes $I_{a,b}$ $=$ $0$, during the same order of Comparing, which is the ultimate refinement of an edge-sequence. If $I_{a,b}$ $=$ $0$, then there is no  possibility of a $K_{C}$, using literals $a, b$, thus no $K_{C}$, using literals $a, b$ and $c$. ie. the bit-change to $c$ in $I_{a,b}$ was subsumed. The only note of interest for the third case is suppose an edge-sequence  $I_{x, y}$  did play a role in the refinement of $I_{a,b}$ with no imposed bit-changes. ie. no $S$. Now suppose that with $S$, $I_{x, y}$ $=$ $0$, during  the same sequence of Comparing. In such a case, at least the  bit-change of $c$ in  $I_{a,b}$ still occurs during the same order of Comparing, because $I_{x, y}$ was just one  edge-sequence that was part of a union of edge-sequences from its $S$-set that refined either  $I_{a,b}$ or it refined another edge-sequence which refined another, and so on, until some union of edge-sequences then refined  $I_{a,b}$.  More generally, the remainder of the union of edge-sequences where $I_{x, y}$ $=$ $0$,  determining an edge-sequence $I$ when Comparing,  will still refine $I$, if the union that included $I_{x, y}$ had. Or, it is the case that the remainder of the union of edge-sequences  determining an edge-sequence $I$, can now refine $I$, because $I_{x, y}$ $=$ $0$. The case being considered for $I_{a,b}$ above, is the former. In summary, refinements do not prevent other refinements, unless one is subsumed by another.\\

\noindent Additionally, it should be clear that if sub edge-sequences of length $p$ cells for some $\mathcal{W}$ with edge-sequences of length $c$ cells, were to be Compared using just the ${p}\choose{2}$\hspace{.1cm} $S$-sets and a refinement $R$  occurred, that increasing the length of those sub edge-sequences to $c$ cells,  prior to the same  order of Comparing, will not prevent refinement $R$, to occur, while processing the same $S$-sets. This of course, assumes that some or all of refinement $R$, was not subsumed while performing the same order of Comparing, due to the addition of $k = c$$- p$ cells. ie. if the additional $k$ cells played no role wrt. refinement $R$. The reason is simply that the refinements were due to the relationships between the first $p$ cells for those same edge-sequences, regardless of the size of $k$. Remember that even a single bit-change always eliminates at least one $K_{C}$ possibility. So, if an edge-sequence $I_{a,b}$ incurs a bit-change in some position for a literal $c$, then there is now, no $K_{C}$ possible, using literals $a, b$ and $c$ together, by definition $2.12$. The theorem to follow establishes that Comparing produces bit-changes in an edge-sequence $I$, for those literals that do not belong to a solution with the endpoints of $I$, where a cell of zero entries eliminates $I$.

\begin{theorem}\label{theorem: $I_{a,b}$ = 0 if}
Let  a  collection of equivalent $S$-sets $\mathcal{X}$  with edge-sequences of length $c$ cells, be the result of Comparing a collection of  $S$-sets $\mathcal{W}$. Then an edge-sequence $I_{x, y}$ from  $\mathcal{X}$, is such that a literal with a $1$ entry in $I_{x, y}$ belongs to at least one $K_{C}$ with $x$ and $y$. \\

\end{theorem}

\noindent Recall that there is no collection of equivalent $S$-sets if unsatisfiability is determined, which is discovered during the first round. \\

\proof The proof is by induction on the number of clauses $c$, and the base case shall be $c$ $=$ $3$. \\

\noindent Let $c$ $=$ $3$.  Observe that the combinations of clause sizes, for three clauses are: \textbf{i}) $3, 3, 3$ \hspace{0.5mm} \textbf{ii}) $3, 3, 2$ \hspace{0.5mm} \textbf{iii}) $3, 2, 2$   \hspace{0.5mm} \textbf{iv}) $3, 3, 1$ \hspace{0.5mm} \textbf{v}) $3, 2, 1$  and \textbf{vi}) $3, 1, 1$.\\

\noindent A $3$-SAT with three clauses  could have as many as $27$ edge-sequences, nine in each $S$-set  constructed or as few as $7$ edge-sequences  for  \textbf{vi}) above. Note that pre-processing would eliminate \textbf{iv}), \textbf{v}) and \textbf{vi}) due to the loner clauses, nor could $27$ edge-sequences be possible without the existence of $\it{pure}$ literals. However, we exclude the pre-processing routine from the proof. \\

\noindent So, after the application of the $LCR$ and $K$-rule to the newly constructed edge-sequences, and any applicable action of the refinement rules, the edge-sequences that remain will have one of the three possible number of $1$ entries:\\

\noindent  \textbf{1}. $\textcolor{blue}{I_{x_{1}, y_{2}}} : ( 1_{x}, 0, 0 \hspace{.15cm}\textcolor{brown}{|} \hspace{.15cm} 1_{y}, 0, 0 \hspace{.15cm}\textcolor{brown}{|} \hspace{.15cm} 1, 0, 0 )$ \\

\noindent \textbf{2}. $\textcolor{blue}{I_{x_{1}, y_{2}}} : ( 1_{x}, 0, 0 \hspace{.15cm}\textcolor{brown}{|} \hspace{.15cm} 1_{y}, 0, 0 \hspace{.15cm}\textcolor{brown}{|} \hspace{.15cm} 1, 1, 0 )$ \\

\noindent \textbf{3}. $\textcolor{blue}{I_{x_{1}, y_{2}}} : ( 1_{x}, 0, 0 \hspace{.15cm}\textcolor{brown}{|} \hspace{.15cm} 1_{y}, 0, 0 \hspace{.15cm}\textcolor{brown}{|} \hspace{.15cm} 1, 1, 1 )$ \\

\noindent $WLOG$, we can assume that the edge-sequence $I_{x, y}$ has its endpoints in the first and second cell. The third cell must have one, two or three $1$ entries, to be $K$-rule compliant. In \textbf{1}, literals $x_{1}$ and $y_{2}$ belong to one $K_{3}$ with some literal having a $1$ entry in the third cell, and in \textbf{2}, $x_{1}$ and $y_{2}$ belong to a $K_{3}$ with each of the two literals in cell three having a $1$ entry. In \textbf{3}, literals $x_{1}$ and $y_{2}$ belong to a $K_{3}$ with each of the three literals in cell three having a $1$ entry.\\

\noindent Claim: If the three $S$-sets having edge-sequences  with $3$ cells, require no more actions from the refinement rules to be taken and are $LCR$, $K$-rule compliant, then the $S$-sets are already equivalent. ie. Only one  \textbf{run} occurs.\\

\noindent Proof: Let literal $x$ be associated to cell $C_{1}$, $y$ to  cell $C_{2}$ and $z$  to  cell $C_{3}$. Suppose $I_{x_{1},y_{2}}$ has a $1$ entry for $z_{3}$. Then let $I_{x_{1},z_{3}}$ have a $1$ entry for $y_{2}$ and $I_{y_{2},z_{3}}$ have a $1$ entry for $x_{1}$. Now, suppose instead that $I_{x_{1},z_{3}}$ has a $0$ entry for $y_{2}$. Then by the action of refinement rule $2$,  $I_{x_{1},y_{2}}$ will have a $0$ entry for $z_{3}$ and  $I_{y_{2},z_{3}}$ will have a $0$ entry for $x_{1}$. Since, the edge-sequences were such that no actions of the refinement rules applied, it can not be the case. Then it is the case that $I_{x_{1}, y_{2}} \cap I_{x_{1}, z_{3}}    \cap  I_{y_{2}, z_{3}} $ $\ne$ $0$, thus, these three edge-sequences belong to a $K_{3}$. Therefore, an edge-sequence $I_{x, y}$ belonging to a collection of three equivalent $S$-sets $\mathcal{X}$, is such that any literal with a $1$ entry in $I_{x, y}$ belongs to at least one $K_{3}$ with $x$ and $y$. \\

\noindent Suppose now that \textbf{c $>$ 3} is an integer for which the statement of
 the theorem   is valid.  We consider cases  \textbf{1},  \textbf{2} and  \textbf{3}. \\

\noindent  \textbf{Case 1}i): Let a collection of $S$-sets $\mathcal{W}$, be for a $3$-SAT $\mathcal{G}$  having $c$$+1$ clauses and a non-singleton literal $x$ that does not belong to any $K_{C+1}$. ie. there is no solution using literal $x$. Let $\mathcal{G}'$ be the $3$-SAT formed with the clauses of $\mathcal{G}$ less one clause: $C_{c+1}$ $=$ $(x, l_{1}, l_{2})$. Let $\mathcal{W}'$ be the collection of $S$-sets for $\mathcal{G}'$. If we apply Comparing to $\mathcal{W}'$, either we determine unsatisfiability in which case, literal $x$ belongs to no solution, or a collection of  equivalent $S$-sets $\mathcal{X}'$, is produced where the hypothesis holds. Suppose there are edge-sequences in $\mathcal{X}'$ of the form: $I_{x, \#}$ where {\footnotesize$\#$} is any literal that forms an edge with $x$.  If $\mathcal{X}'$ contains even one edge-sequence of such a form, then $x$ belongs to a $K_{C}$ $\mathcal{K}$, for $\mathcal{G}'$, so $x$ would also belong to a $K_{C+1}$ for $\mathcal{G}$, using $x$ from the $c$$+1^{th}$ clause. Recall that the literals $x$ would have edges between them and the same literals, outside their respective cells, due to the refinement rules. Therefore, $x$ does not appear in any edge-sequence or vertex-sequence for $\mathcal{G}'$ which implies $x$ does not appear in any edge-sequence or vertex-sequence for $\mathcal{G}$ as well. Because the same refinement would occur or be subsumed by another refinement when Comparing is applied to $\mathcal{W}$ for $\mathcal{G}$, producing a collection of equivalent $S$-sets $\mathcal{X}$. \\

\newpage

\noindent  \textbf{Case 1}ii): Let a collection of $S$-sets $\mathcal{W}$,  be for a $3$-SAT $\mathcal{G}$, having $c$$+1$ clauses and a singleton literal $x$ that does not belong to any $K_{C+1}$. Let $\mathcal{G}'$ be the $3$-SAT formed with the clauses of $\mathcal{G}$ less one clause: $C_{c+1}$ $=$  $(y, l_{1}, l_{2})$, where literal $y$ is a non edge-singleton wrt. $I_{x, y}$ of $\mathcal{W}$. Let $\mathcal{W}'$ be the collection of $S$-sets for $\mathcal{G}'$. If we apply Comparing to $\mathcal{W}'$, either we determine unsatisfiability in which case, literal $x$ belongs to no solution, or a collection of  equivalent $S$-sets $\mathcal{X}'$, is produced where the hypothesis holds. Suppose there is an edge-sequence $I_{x, y}$ from  $\mathcal{X}'$.  If $\mathcal{X}'$ contains edge-sequence $I_{x, y}$, then $x$ belongs to a $K_{C}$ $\mathcal{K}$, for $\mathcal{G}'$, so $x$ would also belong to a $K_{C+1}$ for $\mathcal{G}$, using $y$ from the $c$$+1^{th}$ clause as explained in 1i).  It follows  that $I_{x, y}$ does not appear in $\mathcal{X}'$ which implies it also does not appear in a collection of equivalent $S$-sets $\mathcal{X}$ for $\mathcal{G}$. Again, because the same refinement would occur or be subsumed by another refinement when Comparing is applied to a  collection of $S$-sets $\mathcal{W}$, for $\mathcal{G}$ producing $\mathcal{X}$. If we construct other $3$-SATs with the clauses of $\mathcal{G}$ less one clause with the property of containing at least one literal {\footnotesize$\#_{1}$} that is a non edge-singleton wrt. $I_{x, \#_{1}}$ of $\mathcal{G}$, we see that eventually all edge-sequences with one endpoint associated to $x$ is of the form: $I_{x, \%_{1}}$ where {\footnotesize$\%_{1}$} is an edge-singleton or just a singleton, and every position associated to a non edge-singleton has a zero entry, due to the actions of refinement rule $3$. And by lemma $4.2$, if  $I_{x, \%_{1}}$ is such that its $1$ entries correspond to just edge-singletons or singletons, including $x$ and {\footnotesize$\%_{1}$}, then it belongs to at least one solution. Since $x$ does not belong to a solution, then there is no $I_{x, \%_{1}}$ in $\mathcal{X}$, so it follows that $V_{x}$ $=$ $0$. \\

 \noindent  \textbf{Case 2}i): Let a collection of $S$-sets $\mathcal{W}$, be for a $3$-SAT $\mathcal{G}$,  having $c$$+1$ clauses and an  edge-sequence  $I_{x, y}$ that does not belong to any $K_{C+1}$ where $y$ is a non edge-singleton wrt. $I_{x, y}$. We assume that $x$ and $y$ do belong to different solutions, otherwise we are in case 1. Let $\mathcal{G}'$ be the $3$-SAT formed with the clauses of $\mathcal{G}$ less one clause: $C_{c+1}$ $=$ $(y, l_{1}, l_{2})$. Let $\mathcal{W}'$ be the collection of $S$-sets for $\mathcal{G}'$. If we apply Comparing to $\mathcal{W}'$, either we determine unsatisfiability in which case, $I_{x, y}$  belongs to no solution, or a collection of  equivalent $S$-sets $\mathcal{X}'$, is produced where the hypothesis holds. Suppose $I_{x, y}$ exists in  $\mathcal{X}'$, then $I_{x, y}$   belongs to a $K_{C}$ $\mathcal{K}$, for $\mathcal{W}'$, so $I_{x, y}$  would also belong to a $K_{C+1}$ for $\mathcal{W}$, using $y$ from the $c$$+1^{th}$ clause.   Therefore, $I_{x, y}$ does not appear in $\mathcal{X}'$ which implies $I_{x, y}$  does not appear in $\mathcal{X}$, again, because  the same refinement would occur or be subsumed by another refinement when Comparing is applied to $\mathcal{W}$ for $\mathcal{G}$. \\

\newpage

\noindent \textbf{Case 2}ii): Let a  collection of $S$-sets $\mathcal{W}$,  be for a $3$-SAT $\mathcal{G}$, having $c$$+1$ clauses and an edge-sequence  $I_{x, y}$ that does not belong to any $K_{C+1}$ where $x$ and $y$ are edge-singletons  wrt. $I_{x, y}$. Again, we assume that $x$ and $y$ do belong to different solutions, otherwise we are in case 1. Let $\mathcal{G}'$ be the $3$-SAT formed with the clauses of $\mathcal{G}$ less one clause: $C_{c+1}$ $=$ $(z, l_{1}, l_{2})$, not containing $x$ or $y$, where literal $z$ is a non edge-singleton wrt. $I_{x, y}$. Let  $\mathcal{W}'$ be the collection of $S$-sets for $\mathcal{G}'$. If we apply Comparing to $\mathcal{W}'$, either we determine unsatisfiability in which case, the edge-sequence $I_{x, y}$ belongs to no solution, or a collection of  equivalent $S$-sets $\mathcal{X}'$, is produced where the hypothesis holds. Suppose there is an edge-sequence $I_{x, y}$ from  $\mathcal{X}'$.  If $\mathcal{X}'$ contains edge-sequence $I_{x, y}$ with a $1$ entry for $z$, then $I_{x, y}$  also belongs to a $K_{C+1}$ for $\mathcal{W}$, using the literal $z$ from the $c$$+1^{th}$ clause, contradicting that $I_{x, y}$ $\notin$ any $K_{C+1}$. So if $I_{x, y}$ appears in  $\mathcal{X}'$ it must have a zero entry for literal $z$, and it follows that  the same refinement would occur or be subsumed by another refinement when Comparing is applied to a  collection of $S$-sets $\mathcal{W}$, for $\mathcal{G}$ producing $\mathcal{X}$. If we construct other $3$-SATs with the clauses of $\mathcal{G}$ less one clause not having $x$ or $y$, with the property of containing at least one non edge-singleton wrt. $I_{x, y}$ for $\mathcal{W}$, we see that eventually all that could remain having a $1$ entry in $I_{x, y}$ are  edge-singletons or singletons and every position associated to a non edge-singleton has a zero entry. And by lemma $4.2$, if  $I_{x, y}$ is such that its $1$ entries correspond to only edge-singletons or singletons, then it belongs to at least one solution for $\mathcal{G}$, a contradiction.  Therefore, $I_{x, y}$ does not exist in $\mathcal{X}$. ie. by Comparing $I_{x, y}$ $=$ $0$. \\

\noindent \textbf{Case 3}i): Let an edge-sequence $I_{x, y}$ from a  collection of $S$-sets $\mathcal{W}$,  be for a $3$-SAT $\mathcal{G}$, having $c$$+1$ cells, belong to at least one $K_{C+1}$. First, consider any non edge-singleton literal $s$ wrt. $I_{x, y}$ ($s$ is not $x$ or $y$). Let a collection of equivalent $S$-sets $\mathcal{X}$ having an edge-sequence $I_{x, y}$ be for $\mathcal{G}$. Let $\mathcal{G}'$ be the $3$-SAT formed with the clauses of $\mathcal{G}$ less one clause: $C_{c+1}$ $=$ $(s, l_{1}, l_{2})$. Let $\mathcal{W}'$ be the collection of $S$-sets for $\mathcal{G}'$. If we apply Comparing to $\mathcal{W}'$, a collection of  equivalent $S$-sets $\mathcal{X}'$, is produced where the hypothesis holds. We know that $\mathcal{X}'$ with $I_{x, y}$ will be produced because  $\mathcal{X}$ for $\mathcal{G}$, was produced having edge-sequence $I_{x, y}$. So, if $\mathcal{X}'$ contains edge-sequence $I_{x, y}$ with a $1$ entry for $s$, then $I_{x, y}$  also belongs to a $K_{C+1}$ for $\mathcal{G}$, using the literal $s$ from the $c$$+1^{th}$ clause. If to the contrary, $I_{x, y}$ appeared in  $\mathcal{X}'$ with a zero entry for literal $s$, it would follow that  the same refinement would occur, or be subsumed by another refinement when Comparing is applied to a  collection of $S$-sets $\mathcal{W}$, for $\mathcal{G}$ producing $\mathcal{X}$, where literal $s$ would instead, have a zero entry.

\noindent  Next, consider the cases where both $x$ and $y$ are non edge-singletons each occurring at least twice in $I_{x_{i}, y_{j}}$ or only $x$ is a non edge-singleton which occurs at least three times in $I_{x_{i}, y_{j}}$, where we follow the same process as described in case 1i). For example, suppose  $x$ and $y$ are non edge-singletons each occurring twice in $I_{x_{i}, y_{j}}$. Then, there exists two clauses $C_{m}$ and $C_{n}$ containing $x$ and $y$, respectively. Note that $m$, $n$, $i$ and $j$ are distinct clauses. We shall construct $\mathcal{W}'$ $=$ $\mathcal{W} - \{ C_{m}\}$ and $\mathcal{W}''$ $=$ $\mathcal{W} - \{ C_{n}\}$ and follow the process as outlined in 1i). For  $\mathcal{W}'$ $=$ $\mathcal{W} - \{ C_{m}\}$, we are able to determine the  entries for all cells except cells $m$ and $i$, when examining $I_{x_{i}, y_{j}}$ and $I_{x_{i}, y_{n}}$. And for $\mathcal{W}''$ $=$ $\mathcal{W} - \{ C_{n}\}$,  we are able to determine the  entries for all cells except cells $j$ and $n$, when examining $I_{x_{i}, y_{j}}$ and $I_{x_{m}, y_{j}}$. However, all together, every cell had its  entries determined at size $c$, where the hypothesis holds, which implies the entries are also known at size $c$$+1$, for these edge-sequences. Now, suppose that $x$ is a non edge-singleton which occurs three times in $I_{x_{i}, y_{j}}$ and $y$ is an edge-singleton wrt. $I_{x_{i}, y_{j}}$. Then, there exists three distinct clauses $C_{m}$, $C_{n}$ and $C_{i}$ containing $x$. We shall construct $\mathcal{W}'$ $=$ $\mathcal{W} - \{ C_{m}\}$ and $\mathcal{W}''$ $=$ $\mathcal{W} - \{ C_{i}\}$ then follow the process as outlined in 1i).  For  $\mathcal{W}'$ $=$ $\mathcal{W} - \{ C_{m}\}$, we are able to determine the  entries for all cells except cell $m$, when examining $I_{x_{i}, y_{j}}$ and $I_{x_{n}, y_{j}}$. And for $\mathcal{W}''$ $=$ $\mathcal{W} - \{ C_{i}\}$,  we are able to determine the  entries for all cells except cell $i$, when examining $I_{x_{n}, y_{j}}$ and $I_{x_{m}, y_{j}}$. Again, all together, every cell had its  entries determined at size $c$, where the hypothesis holds, so it follows that the entries are also known at size $c$$+1$ for these edge-sequences. \\

\noindent \textbf{The  $I_{x, x}$ edge-sequences}: Observe that the edge-sequences whose endpoints correspond to the same literal such as $I_{x, x}$, are never removed by Comparing, if $x$ belongs to at least one solution. Moreover, each instance of $x$ from any two clauses, $x_{i}$ and $x_{j}$, will have the same vertex-sequences outside cells $i$ and $j$. So, if a literal $z$, with a $1$ entry in $I_{x, x}$ belongs to at least one solution with $x$, then the edge-sequence $I_{x, z}$ must exist. Conversely, if  Comparing produces $I_{x, z}$ $=$ $0$, then there will be a zero entry corresponding to $z$ in $I_{x, x}$ due to the actions of refinement rule 3. \\

\noindent \textbf{Case 3}ii): Let a  collection of $S$-sets $\mathcal{W}$  for a $3$-SAT $\mathcal{G}$, with $c$$+1$ clauses have an edge-sequence $I_{x, y}$, where $x$, $y$ and a literal $z$ are  edge-singletons  wrt. $I_{x, y}$ such that there is no $K_{C+1}$ using  $x$, $y$ and $z$ together. We assume that $I_{x, y}$ belongs to at least one $K_{C+1}$, otherwise we are in case 2ii). So, it must be established that Comparing will produce a zero entry for $z$ in $I_{x, y}$. \\

\noindent There are two special cases of note.  \textbf{1})  Let $\mathcal{G}'$ be a $3$-SAT formed with the clauses of $\mathcal{G}$ less one clause not containing $x$, $y$ or $z$, where the collection of $S$-sets $\mathcal{W}'$ for $\mathcal{G}'$, does not possess a $K_{C}$ using $x$, $y$ and $z$ together. Then Comparing  $\mathcal{W}'$ will produce  $I_{x, y}$ with a zero entry for $z$, and by refinement rule $2$,  $I_{x, z}$ and  $I_{z, y}$ incur bit-changes for $y$ and $x$, respectively, and as explained in cases above, this implies the same outcome wrt.  $I_{x, y}$, $I_{x, z}$ and  $I_{y, z}$ of $\mathcal{W}$ for $\mathcal{G}$.  \textbf{2})  Since we do not exclude a $3$-SAT with one or more $\it{pure}$ literals, $\it{quantums}$ or even duplicate clauses, then if one of these should exist outside the clauses containing $x$, $y$ and $z$, simply process the collection of $S$-sets  $\mathcal{W}'$ for the $3$-SAT $\mathcal{G}'$ formed with the clauses of $\mathcal{G}$ less a clause that's a duplicate,  a $\it{quantum}$ or contains a $\it{pure}$. It must be the case that Comparing $\mathcal{W}'$  produces a bit-change for $z$ in $I_{x, y}$ of $\mathcal{W}'$, and again, it will be the same outcome for  $I_{x, y}$ of $\mathcal{W}$, having $c$$+1$ cells. \\

\noindent Next, a description is provided regarding the mechanics of edge-sequence intersections during Comparing resulting in a bit-change for an edge-sequence that is being \textbf{determined}. Consider a  collection of $S$-sets $\mathcal{W}$  for a $3$-SAT $\mathcal{G}$, with $c$ clauses, having an edge-sequence $I_{d, f}$ being determined by the edge-sequences from an $S$-set  $\mathcal{E}$. If there are zero entries for  either $d$ or $f$ in the three (at most), edge-sequences  with a common endpoint $e$ say, in $\mathcal{E}$,  then determining  $I_{d, f}$ by $\mathcal{E}$, results in a zero entry for $e$ in $I_{d, f}$. It is possible that at least one of the three edge-sequences had a zero intersection with  $I_{d, f}$ due to other cells and not because of zero entries for $d$ or $f$. Note that if just one of  $d$ or $f$,  had all the zero entries  in the three edge-sequences with common endpoint $e$, then $I_{d, e}$ or $I_{f, e}$ would be zero, again resulting in a zero entry for $e$ in $I_{d, f}$. The two other possible ways that literal $e$ incurs a bit-change in  $I_{d, f}$  would be if $e$ was not an endpoint to any edge-sequence in  $\mathcal{E}$, but did have a zero entry in every edge-sequence in $\mathcal{E}$ which had a non-zero intersection with $I_{d, f}$. Or, if its negation $-e$, becomes a loner cell literal, so $e$ incurs a bit-change to ensure $LCR$ and $K$-rule compliancy of $I_{d, f}$. \\

\noindent  \textbf{The literal triples}: Every collection of $S$-sets $\mathcal{W}$, with edge-sequences of length $c$,  for a $3$-SAT $\mathcal{G}$, have a set $\mathcal{L}$, consisting of all triples of literals from three distinct cells, for which no two are each other's negation. There is a partition of  $\mathcal{L}$ by two sub-sets $\mathcal{M}$ and $\mathcal{N}$, where $\mathcal{M}$ are those that do belong to a $K_{C}$ together, and $\mathcal{N}$ are those that do not belong to a $K_{C}$ together. If  $\mathcal{N}$ is the null set, then $\mathcal{L}$ $=$  $\mathcal{M}$. If $\mathcal{N}$ is not null, then the triples partition again into two sets $\mathcal{A}$ and $\mathcal{B}$. Set $\mathcal{A}$ are those triples which exist due to $LCR$ and $K$-rule compliancy when constructing the edge-sequences for $\mathcal{W}$. More precisely, if three literals $a$, $b$ and $c$  belong to clause $C_{i}$, so cell $C_{i}$ as well, having negations in three distinct clauses, then edge-sequences $I_{-a, -b}$, $I_{-a, -c}$ and $I_{-b, -c}$ have bit-changes to $-c$, $-b$ and $-a$ respectively, to be $LCR$ compliant, which makes the triple $-a$, $-b$ and $-c$ an $\mathcal{A}$ set triple. Note well that membership of a triple to set $\mathcal{A}$ implies Comparing did not produce this triple, thus they exist before any Comparing is done. Observe that in the case where a clause for $\mathcal{G}$ has just two literals, say  $C_{i}$ $=$ $(a, b)$, then the triple: $-a$, $-b$ and {\footnotesize$k$} is an $\mathcal{A}$ set triple, where {\footnotesize$k$} is any literal that can form edges with $-a$ and $-b$, with $I_{-a, k}$ and $I_{-b, k}$  incurring bit-changes to $-b$ and $-a$ respectively, but $I_{-a, -b}$ $=$ $0$.  To see that an $\mathcal{A}$ set triple must be from three distinct clauses, let a $3$-SAT $\mathcal{G}$, have three clauses: $(a, b, c)$, $(-a, -b, l_{1})$ and $(-c, l_{2}, l_{3})$, where  $a$, $b$, $c$, $-a$, $-b$ and $-c$, are all singletons. Note that edge-sequences $I_{-a, -c}$ and $I_{-b, -c}$  do not incur bit-changes to $-b$ and $-a$ respectively, to be $LCR$ compliant, as they are in the endpoint cells, and $I_{-a, -b}$ does not exist. So, the triple $-a$, $-b$ and $-c$ is not an $\mathcal{A}$ set triple because no refinements for $LCR$ compliancy occurred. Now, suppose a $3$-SAT $\mathcal{G}$, has two clauses: $(a, b, c)$ and $(-a, -b, -c)$ where  $a$, $b$, $c$, $-a$, $-b$ and $-c$, are all singletons. Then no edge-sequences can be formed, thus no bit-change occurred, so the triple $-a$, $-b$ and $-c$ is not an $\mathcal{A}$ set triple. If prior to any Comparing, we have that $I_{x, y}$ $=$ $0$,  then $x$, $y$ and some other literal are $\mathcal{A}$ set triples. If prior to any Comparing, we have that $V_{x}$ $=$ $0$, then every pair of endpoints and $x$ are $\mathcal{A}$ set triples. Set $\mathcal{B}$ are those triples $x$, $y$ and $z$, that only by Comparing have edge-sequences $I_{x, y}$, $I_{x, z}$ and $I_{y, z}$ with bit-changes to $z$, $y$ and $x$ respectively. If by Comparing, an edge-sequence $I_{x, y}$ or vertex-sequence $V_{x}$, equals zero, then bit-changes only occur to  $x$ and $y$ or $x$ respectively, in the edge and vertex sequences that remain. To have a set $\mathcal{B}$ requires the existence of a set $\mathcal{A}$. That is, without a set $\mathcal{A}$, there is no set $\mathcal{B}$, which means every triple of $\mathcal{W}$ is in $\mathcal{M}$. Most importantly, it must be the case that the $\mathcal{A}$ set triples which exist before any Comparing, determine which triples will become $\mathcal{B}$ set triples, by Comparing. ie. the $\mathcal{A}$ set triples in $\mathcal{W}$ determine prior to Comparing, which three literals can belong to a solution together, where the ones who can not become  $\mathcal{B}$ set triples. Now observe that the negations of the literals from each clause of size $3$, for a $3$-SAT are a collection of $\mathcal{A}$ set triples, if no $\it{pures}$ or $\it{quantums}$ exist and no edge or vertex sequence became zero prior to Comparing. Suppose we have a $3$-SAT $\mathcal{G}$ with $k$ clauses and its collection of $S$-sets $\mathcal{W}$ contains an edge-sequence $I_{x, y}$ where $x$, $y$ and $z$ are edge-singletons, in clause/cells $s$, $t$ and $r$, respectively. Let there be no solution using $x$, $y$ and $z$ together. If we are not in special case \textbf{1}) wrt. $I_{x, y}$, then it's the case that forming a $3$-SAT by removing a clause $q$ which is not clause $s$, $t$ or $r$, results in a solution that uses $x$, $y$ and $z$ together. Moreover, every solution using $x$, $y$ and $z$ together, must also use the negations of the literals of $q$. This is true for any clause removed that's not $s$, $t$ or $r$. So, the solutions are distinct from one another assuming no clause is duplicated. Let the set $\mathcal{P}_{i}$ be a collection of $\mathcal{A}$ set triples with the property that no literal and its negation exists among them. We assert that the mechanics of edge-sequence intersections during Comparing resulting in a bit-change for $z$ in $I_{x, y}$ (or $x$ in $I_{y, z}$ or $y$ in $I_{x, z}$), is due solely to the existence of at least one  $\mathcal{P}_{i}$ wrt. the literal triple: $x$, $y$ and $z$, in $\mathcal{W}$. Suppose it's true in general for every size $k$ $\le$ $c$. Then $WLOG$, let a clause: $C_{w}$ $=$ $(a, b, c)$ be for $\mathcal{G}$ with $c$$+1$ clauses, where the negations for the literals of cell $w$ are in three distinct cells which are not the cells $r$, $s$ and $t$. Note that $-x$, $-y$ or $-z$ (all necessarily exist to be a non-trivial case), could be one of $a$, $b$ or $c$. Then let $\mathcal{Y}$ be the collection of $S$-sets produced by removing all $S$-sets of the form: $S_{w, \%_{2}}$ where {\footnotesize$\%_{2}$} is any cell index not $w$, from  $\mathcal{W}$ of $\mathcal{G}$, as well as removing cell $w$ from  all the edge-sequences in the remaining $S$-sets. Observe that there is no $K_{C}$ using $x$, $y$ and $z$ together, in $\mathcal{Y}$. Since a set $\mathcal{P}_{i}$ wrt.  the literal triple: $x$, $y$ and $z$ exists in $\mathcal{Y}$, which includes the $\mathcal{A}$ set triple: $-a$, $-b$ and $-c$, the relationship necessary to produce bit-changes to $z$, $y$ and $x$  in $I_{x, y}$, $I_{x, z}$ and $I_{y, z}$, respectively, by Comparing $\mathcal{Y}$, exists. The implication for  all $3$-SATs  having a literal triple requiring another literal triple in order to be part of a solution, is that if the latter triple is turned into an $\mathcal{A}$ set triple (\textbf{imposed} bit-changes), then the dependent triple must become a $\mathcal{B}$ set triple, by Comparing. Note that the pair of triples could be co-dependent, in which case either could be the $\mathcal{A}$ set triple while the other the  $\mathcal{B}$ set triple. Of course, it must be established that a maximum sized  $\mathcal{P}_{i}$ wrt.  the triple $x$, $y$ and $z$ at $c$$+1$, results  in bit-changes to $z$, $y$ and $x$  in $I_{x, y}$, $I_{x, z}$ and $I_{y, z}$, respectively, by Comparing. So we define a $\it{truncated}$ sequence to be an edge-sequence less one or both of its endpoint cells. Recall that, a (sub) edge-sequence has one or more cells removed but neither cell containing the  endpoints are removed. We also define a $\it{truncated}$ $S$-set to be an $S$-set with at most 9 $\it{truncated}$ sequences. In particular, we consider the $\it{truncated}$ sequences: $I_{\textcolor{red}{z}, \#_{7}}$, $I_{\textcolor{red}{x}, \#_{2}}$ and $I_{\textcolor{red}{y}, \#_{3}}$. Next we take $\mathcal{W}$ -  \{$S_{r, \#_{4}}$\},  $\mathcal{W}$ -  \{$S_{s, \#_{5}}$\} and  $\mathcal{W}$ -  \{$S_{t, \#_{6}}$\} to produce the collections of equivalent $S$-sets: $\mathcal{U}'$, $\mathcal{U}''$ and $\mathcal{U}'''$, respectively, by Comparing. Observe that the $S$-sets have edge-sequences of length $c$$+1$ cells and assume the bit-changes of interest did not occur in the $r$, $s$ or $t$ cells. An intersection between an edge-sequence $I$, from one of these $S$-sets and a $\it{truncated}$ sequence will be zero, if $I$ has a zero entry for the truncated endpoint. Otherwise, assume a $1$ entry for the truncated endpoint and zero entries for the other two positions, for the intersection. And the $\it{truncated}$ sequences will have the same truncated endpoint cell for the Comparing to be done, so an intersection between them will be zero if they don't have the same  truncated endpoint. Now we Compare $S_{\textcolor{red}{r}, \#_{4}}$ with $\mathcal{U}'$, $S_{\textcolor{red}{s}, \#_{5}}$ with $\mathcal{U}''$ and $S_{\textcolor{red}{t}, \#_{6}}$ with $\mathcal{U}'''$. We also assume that at no time will an edge-pure literal be produced that's not in cells $r$, $s$ or $t$ of the truncated edge-sequences: $I_{\textcolor{red}{z}, x}$, $I_{\textcolor{red}{z}, y}$, $I_{\textcolor{red}{x}, y}$, $I_{\textcolor{red}{x}, z}$, $I_{\textcolor{red}{y}, x}$ or $I_{\textcolor{red}{y}, z}$. Else, the literal triple $x$, $y$ and $z$ can be determined with just $c$ cells simply by  removing a cell with such an edge-pure literal from the $S$-sets and apply Comparing. We further assume that no refinement of $\mathcal{U}'$, $\mathcal{U}''$ and $\mathcal{U}'''$ due to the $\it{truncated}$ sequences of $S_{\textcolor{red}{r}, \#_{4}}$, $S_{\textcolor{red}{s}, \#_{5}}$ and $S_{\textcolor{red}{t}, \#_{6}}$ occurred which led to $x$, $y$ and $z$ becoming a $\mathcal{B}$ set triple. Since all bit-changes by Comparing occur outside the endpoint cells (recall that the zero entries for all $\mathcal{A}$ set triples using truncated endpoints exist in $\mathcal{U}'$, $\mathcal{U}''$ and $\mathcal{U}'''$), then the result will be a bit-change to at least one of the literals $x$, $y$ or $z$ in one of the $\it{truncated}$ sequences of interest, which are  treated as vertex-sequences with $c$ cells,  and $\mathcal{P}_{i}$ is wrt. their literal doubles: ($x$, $y$), ($y$, $z$) and ($x$, $z$), which will become $\mathcal{B}$ set doubles. Note that if  a position for the negation of a truncated endpoint exists (having a zero entry), it can be part of a $\mathcal{P}_{i}$, as a literal of an $\mathcal{A}$ set double (with the other endpoint of the $\it{truncated}$ sequence), even if the truncated endpoint belongs to an $\mathcal{A}$ set for $\mathcal{P}_{i}$, because any solution represented in a $\it{truncated}$ sequence can always include the literal associated to the truncated endpoint. Another way to see that the bit-changes of interest will occur just by the assertion holding at size $c$, is to replace all the $S$-sets  with an index of $s$ (or all with an index of $t$),  in $\mathcal{U}'$, with the $S$-sets: $S_{\textcolor{red}{r}, \#_{4}}$ and apply Comparing again. The result will be a zero entry for $y$ and $x$ in  $I_{\textcolor{red}{z}, x}$ and $I_{\textcolor{red}{z}, y}$, respectively, even though they have a vertex-sequence structure wrt. the $c$ cells of truncation. Note that if a  $K_{C}$ uses $x$ and $y$ together, it must necessarily allow including the negation of the truncated endpoint which is not possible in $I_{\textcolor{red}{z}, x}$ or $I_{\textcolor{red}{z}, y}$. Similarly, $\mathcal{U}''$ or $\mathcal{U}'''$ with their corresponding $S$-sets could have been chosen. Therefore, Comparing produces zero entries in edge-sequences, for literals that do not belong to a $K_{C+1}$ with the endpoints of those edge-sequences. We conclude that the inductive statement is valid at $c$$+1$  which completes the proof of the induction step, and so the result follows.

\endproof

\begin{lemma}\label{lemma: algo is polynomial}
The algorithm applied to any $3$-SAT  $\mathcal{G}$, determines satisfiability in polynomial time.

\end{lemma}

\proof By theorem $4.1$, a $3$-SAT is known to be satisfiable at the completion of round $1$ without necessarily constructing a solution. To make the complexity analysis straightforward for determining satisfiability, we take into account no efficiencies. Further, the worst case scenario assumes redundancies that no $3$-SAT would have throughout the processing. Let $n$ be the sum of the sizes for all $c$ clauses, that remained after pre-processing, for a given $3$-SAT  $\mathcal{G}$. Then $c$ $\le$ $n$ $\le$ $3c$. We shall assume that it always took an entire \textbf{run} to produce just one bit-change. We shall further assume that to determine satisfiability, it was required that every position in every edge-sequence incurred a bit-change, which of course, would never occur.\\

\noindent The number of edges is $\le$ ${n}\choose{2}$. Thus, the total number of positions:
 \[ {{{n}\choose{2}} n} = \bigg(\frac{(n)(n-1)}{2}\bigg) n = \bigg(\frac{n^{2}-n}{2}\bigg)n  < n^{3} \] \\


\noindent Recall, that the number of $S$-sets = ${c}\choose{2}$. And a \textbf{run} would be:

\[   {{{c}\choose{2}}\choose{2}} =    \frac{ {{c}\choose{2}}^{2} - {{c}\choose{2}} }{2}   = \frac{ {\bigg(\frac{c^{2} - c}{2}\bigg)}^{2} - {\bigg(\frac{c^{2} - c }{2}\bigg) } }{2}  < c^{4} \le n^{4}  \]

\vskip 3mm
\noindent Now, we need to determine the complexity for a comparison between two $S$-sets. First, we  point out that if only one bit-change was supposed to have occurred for each \textbf{run}, implies that for all Compares of two $S$-sets, $S_{i, j}$ and $S_{k, l}$  just $S_{i,j}$ $\overset{1}{\underset{2}{\rightleftharpoons}}$ $S_{k,l}$ occurred. ie. every edge in $S_{i,j}$ was \textbf{determined} by the edges of $S_{k,l}$  and  every edge in $S_{k,l}$ was \textbf{determined} by the edges of $S_{i,j}$ and no more was done between those two $S$-sets. Of course, one Comparing did have a bit-change occur. In this case, at most, one more determination of every edge in $S_{i,j}$ was done with the edges of $S_{k,l}$ again, because the bit-change could have occurred during: $S_{i,j}$ $\overset{2}{\leftharpoondown}$ $S_{k,l}$.  The worst case for Comparing in one direction: $S_{i,j}$ $\rightharpoonup$ $S_{k,l}$  would be that every edge-sequence in $S_{k,l}$ have $1$ entries for both endpoints of every edge in $S_{i,j}$. If we assume $2n$ steps are required to compare position by position, two sequences of length $n$, then all of the intersections and unions needed to process one edge is $36(n)$, which includes determining the particular refinement, if any. And if all $9$ edges are the same scenario, then it takes $324(n)$ steps for one direction. Assuming we have the same scenario comparing in the other direction, it would be a total of $648(n)$ steps. Let $m$ be the constant for the work $m(n)$ to do $LCR$ compliancy. $LCR$ is clearly linear, since the work is a look-up and bit-changes for the negations of loner cell literals. $K$-rule compliancy is known during the $2n$ steps to compare positions, as is the discovery of  new loner cells.  Now, let $M$ be the constant $648$ + $m$, where the $648$, is for the $648(n)$ steps for all Compares less one which might have used an additional $324(n)$ steps.  Then the work for comparing two $S$-sets with no bit-change is $M(n)$. The additional work for: $S_{i,j}$ $\overset{3}{\rightharpoonup}$ $S_{k,l}$  would be for an entire \textbf{run}. However, we could choose to make $M$ equal to $648$ + $324$ + $m$, which says every Comparing of two $S$-sets, did $S_{i,j}$ $\overset{3}{\rightharpoonup}$ $S_{k,l}$. It is important to note that the algorithm would produce the same result if only $S_{i,j}$ $\overset{1}{\underset{2}{\rightleftharpoons}}$ $S_{k,l}$  was done regardless if  bit-changes occurred. The only computational difference is that it may cause more \textbf{runs} to occur, because no follow up Compares were done in the current \textbf{run}. However, if another bit-change would have occurred say doing:  $S_{i,j}$ $\overset{3}{\rightharpoonup}$ $S_{k,l}$  then that scenario awaits to be done in the next \textbf{run}, as explained in section $3$.  But with respect to the worst case, this is irrelevant, because we are assuming that the maximum number of \textbf{runs} possible, were done.\\


\noindent Finally, taking the product for: the work of comparing two $S$-sets ($Mn$), with the number of $S$-sets compared per \textbf{run} ($n^{4}$), and the most \textbf{runs} possible ($n^{3}$), is the worst case. This gives a complexity of $\mathcal{O}(Mn^{8})$, in Big O notation.

\endproof

\noindent It's straightforward to show that a solution can be constructed with complexity $\mathcal{O}(Mn^{9})$. Just take any edge-sequence $I_{x, y}$ at the end of round $1$, remove the cells containing an $x$, $y$ and a third literal $z$, who had a $1$ entry in $I_{x, y}$. Then make all positions for $-z$ a zero. Next, reconstruct a new $3$-SAT with clauses matching the $1$ entries in the remaining cells in  $I_{x, y}$ and apply the algorithm again. There will be equivalent $S$-sets where again we can select an edge-sequence and repeat the above. This can only be done at most $\lfloor\frac{c}{3}\rfloor$   times. Thus, we have at most $(n^8) (\frac{c}{3})$ $<$ $n^9$. However, it's been established that an algorithm can construct a solution, also with complexity $\mathcal{O}(Mn^{8})$. \\

\noindent The most important empirical observation with respect to efficiency, is after processing many $3$-SATs considered to be hard instances, our original version which had no efficiencies including the actions for the refinement rules,  never approached $n^{7}$ to construct a complete solution,  where $n$ is the number of clauses! Thus, we expect any version with one or more non-trivial efficiencies added, to do better than the bound observed empirically. That is why we expect that efficient versions will be practical with respect to today's hardware. Still, it would be reasonable to expect a bound of $n^{6}$ on average, without efficiencies. However, there are numerous schemes that allow processing of a round to be much less work, than  the work done by the algorithm presented. Of course, it had to be established first,  that a general method exists, namely, the Comparing of the $S$-sets until equivalency is reached.   \\


\section{Final comments}

\noindent Most of our research was devoted to finding efficiencies for both general purpose solving and for specific classes of SATs. We have found dozens so far, and it's  clear many more can be developed. We observed empirically, that with respect to constructing a solution,  the combined work done in subsequent rounds was only a fraction of the work done in round $1$. This observation is not why the worst case bound remains unchanged for constructing a solution. One powerful efficiency that was developed, constructs a solution while dispensing with completing rounds altogether, so in particular round $1$. \\

\noindent The algorithm described herein does not make use of all the information that is determined while Comparing $S$-sets, that if used, would only increase the space complexity by adding at most $n^{3}$ for lookup tables, while reducing the number of \textbf{runs} and dramatically reducing the number of edge-sequence intersections performed. We are assuming that always doing a lookup is less work than performing every intersection when $n$ is sufficiently large. Additionally, well known efficiency schemes could take on new relevance with respect to processing edge-sequences,  by providing analysis at relatively no cost.  They will also become  non-trivial efficiency schemes for many classes of $3$-SATs. \\

\newpage

\noindent We have developed a way for parallel implementation that requires no communication between the nodes that is not the master node,  and at that, it only requires passing information to update the sequences, at specific times.\\

\noindent There is a natural generalization of the algorithm described in this paper for SAT. This could be exploited for efficiency purposes, by extracting information at chosen costs, for Comparing a SAT's corresponding $3$-SAT. It is the case that Comparing for SAT becomes more expensive as clause size increases relative to just converting to $3$-SAT. \\

\noindent A lot of time was dedicated in the last few years for  pre-processing analysis, to determine the best initial choices for Comparing $S$-sets. We experimented by choosing certain $S$-sets to Compare first,  which in many cases had a dramatic effect on efficiency, and it was these observations that inspired one of our most powerful efficiency schemes. \\

\noindent Lastly, note that code with a few implementation modifications employed, has a complexity of $\Omega$$(n^{3})$ for $3$-SATs of semi-primes. \\


\newpage

\begin{center}

  \Large {\textbf{Example re-visited}}

\end{center}


\noindent The $3$-SAT below is  Figure $1$ with all its edges shown, prior to any application of $LCR$ and $K$-rule to the corresponding edge-sequences. The blue edges correspond to the brown edges for the $K_{5}$ depicted in Figure $1$.  The edges in red (which will fail to be $LCR$ and $K$-rule compliant), are those between the vertex associated to literal $a$ and a vertex associated to some other literal that is not literal $-a$. Note that using the pre-processing routine would have precluded the construction of edge and vertex sequences because a solution will be found during this process. First, clause $C_{1}$ contains two $\it{pure}$ literals, namely $x$ and $y$, so the clause is removed. Then with respect to the four remaining clauses, literal $-a$ is $\it{pure}$, and appears in each clause, so all the clauses are now removed. If the version that constructs a solution is used, the output will be: $x$, $-a$ assuming the ordering: $x$ $<$ $y$ in $C_{1}$. \\


\begin{center}
\includegraphics[width=4.5in,height=4.5in]{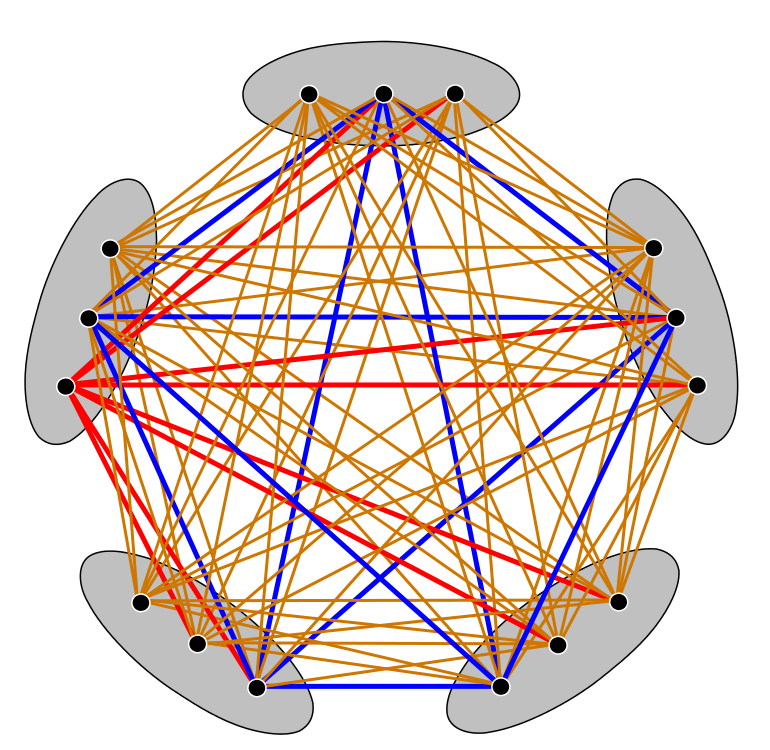}
\end{center}

\noindent  The $3$-SAT below is again  Figure $1$, where Comparing was applied to a collection of 10 $S$-sets with $LCR$ and $K$-rule compliant edges-sequences which produced a collection of 10 equivalent $S$-sets. The result is that every edge depicted below belongs to at least one $K_{5}$.  \\

\begin{center}
\includegraphics[width=4.5in,height=4.5in]{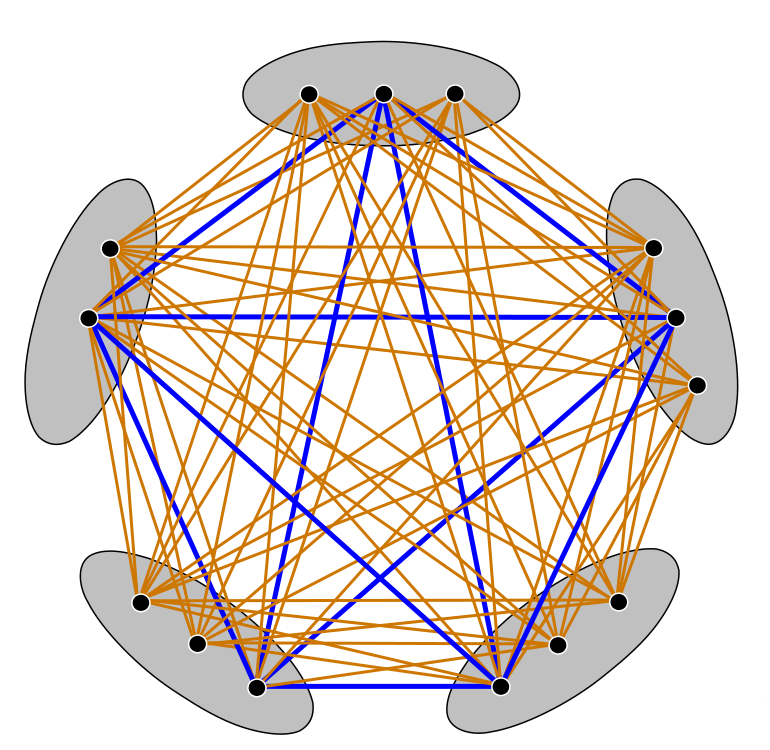}
\end{center}

\vskip 20mm

\begin{center}
\Huge {\textbf{DON'T PANIC}}

\end{center}


\begin {thebibliography}{999}

 \bibitem{cook71}
 Cook, Stephen \\
\emph {The complexity of theorem proving procedures}. \\
Proceedings of the Third Annual ACM Symposium on Theory of Computing.\\ pp. $151$-$158$,
1971.

\bibitem{karp72}
 Karp, Richard M.\\
\emph{Reducibility Among Combinatorial Problems}. \\
Complexity of Computer Computations. New York: Plenum.\\ pp. $85$-$103$,
1972.


\bibitem{agrawal04}

Agrawal, Manindra; Kayal, Neeraj; Saxena, Nitin. \\
\emph{PRIMES is in P}. \\
Annals of Mathematics. \\ pp. $781$-$793$,  2004.




\end{thebibliography}
\end{document}